\documentclass[
    aps,
    prd,
    reprint,
    superscriptaddress,
    nolongbibliography,
    nobibnotes,
    nofootinbib,
]{revtex4-2}

\usepackage[
    colorlinks=true,
    linkcolor=cyan,
    citecolor=cyan,
    filecolor=cyan,
    urlcolor=cyan,
    pdfusetitle,
]{hyperref}

\usepackage[dvipsnames]{xcolor}
\usepackage{amssymb}
\usepackage{amsmath} 
\usepackage{physics}
\usepackage{mathtools}
\usepackage{orcidlink}
\usepackage[normalem]{ulem}
\usepackage{array}

\usepackage[caption=false]{subfig}

\usepackage{acronym}
\newacro{SNR}[SNR]{signal-to-noise ratio}
\newacro{FAR}[FAR]{false-alarm rate}
\newacro{LVK}[LVK]{LIGO--Virgo--KAGRA}
\newacro{GW}[GW]{gravitational wave}
\newacro{BH}{black hole}
\newacro{BBH}{binary black hole}
\newacro{CI}{credible interval}
\newacro{O4}{fourth observing run}
\newacro{KL}{Kullback--Leibler}
\newacro{PPD}{posterior population distribution}
\newacro{ML}{maximum likelihood}
\newacro{GWTC-3}{third gravitational-wave transient catalog}
\newacro{GWTC-4}{fourth gravitational-wave transient catalog}

\newcommand{\ligo}{\affiliation{LIGO Laboratory, Massachusetts Institute of Technology, Cambridge, MA 02139, USA}}
\newcommand{\mki}{\affiliation{Kavli Institute for Astrophysics and Space Research, Massachusetts Institute of Technology, Cambridge, MA 02139, USA}}
\renewcommand{\mit}{\affiliation{Department of Physics, Massachusetts Institute of Technology, Cambridge, MA 02139, USA}}

\def\lvar{\ensuremath{\mathcal{V}}}
\def\dvar{\ensuremath{v_{\rm det}}}
\def\pevar{\ensuremath{v_i}}

\def\catinj{\ensuremath{\mathcal{D}_{\mathrm{inj}}}}
\def\neff{\ensuremath{N_{\rm eff}}}
\def\ndet{\ensuremath{N_{\rm det}}}
\def\ninj{\ensuremath{N_{\rm inj}}}
\def\pdet{\ensuremath{\alpha_{\rm det}}}
\def\nobs{\ensuremath{N_{\rm obs}}}
\def\npe{\ensuremath{N_{\rm PE}}}
\def\mpli{\ensuremath{\lambda_m}} 
\def\mgbr{\ensuremath{\xi_m}} 
\def\ldraw{\ensuremath{\Lambda^{\rm inj}}}
\def\ltrue{\ensuremath{\Lambda^{\rm true}}}
\def\nmin{\ensuremath{N_{\rm inj}^{\min{}}}}
\def\vpe{\ensuremath{\lvar_{\rm PE}}}
\def\vdet{\ensuremath{\lvar_{\rm det}}}

\def\relunc{\ensuremath{\zeta}}
\def\fhope{\ensuremath{\gamma}}

\graphicspath {{figures/}}

\begin{document}

\title{
Reducing systematic uncertainties in gravitational-wave population analyses by increasing the detection threshold
}

\author{Noah E. Wolfe\,\orcidlink{0000-0003-2540-3845}}
\email{newolfe@mit.edu}
\ligo\mki\mit
\author{Matthew Mould\,\orcidlink{0000-0001-5460-2910}}
\ligo\mki\mit
\affiliation{School of Mathematical Sciences, University of Nottingham, Nottingham, NG7 2RD, United Kingdom}
\author{Jack Heinzel\,\orcidlink{0000-0002-5794-821X}}
\ligo\mki\mit
\author{Salvatore Vitale\,\orcidlink{0000-0003-2700-0767}}
\ligo\mki\mit

\date{\today}

\begin{abstract}
From catalogs of gravitational-wave transients, the population-level properties of their sources and the formation channels of merging compact binaries can be constrained.
However, astrophysical conclusions can be biased by misspecification or misestimation of the population likelihood.
Despite detection thresholds on the \ac{FAR} or \ac{SNR}, the current catalog is likely contaminated by noise transients.
Further, computing the population likelihood becomes less accurate as the catalog grows.
Current methods to address these challenges often scale poorly with the number of events and potentially become infeasible for future catalogs.
Here, we evaluate a simple remedy: increasing the significance threshold for including events in population analyses.
To determine the efficacy of this approach, we analyze simulated catalogs of up to 1600 gravitational-wave signals from black-hole mergers using full Bayesian parameter estimation with current detector sensitivities.
We show that the growth in statistical uncertainty about the black-hole population, as we analyze fewer events but with higher \ac{SNR}, depends on the source parameters of interest.
When the \ac{SNR} threshold is raised from 11 to 15---reducing our catalog size by two--thirds---we find that statistical uncertainties on the mass distribution only grow by a few 10\% and constraints on the spin distribution are essentially unchanged; meanwhile, uncertainties on the high-redshift cosmic merger rate more than double.
Simultaneously, numerical uncertainty in the estimate of the population likelihood more than halves, allowing us to ensure unbiased inference without additional computational expense.
Our results demonstrate that focusing on higher-significance events is an effective way to facilitate robust astrophysical inference with growing gravitational-wave catalogs.

\end{abstract}

\acresetall{}

\maketitle

\section{Introduction}

Since the first detection of gravitational waves (GWs) \acused{GW} \cite{LIGOScientific:2016aoc}, the catalog of observed \ac{GW} transients has grown to 90 candidates \cite{LIGOScientific:2018mvr, LIGOScientific:2021usb, KAGRA:2021vkt} by the end of the third observing run of the \ac{LVK} \cite{LIGOScientific:2014pky, VIRGO:2014yos, KAGRA:2020tym} collaboration and 218 candidates through the first part of the \ac{O4} \cite{LIGOScientific:2025slb}.
All of these events have been identified as compact-object mergers involving black holes (BHs) \acused{BH} or neutron stars.

Combining multiple events together and analyzing them jointly can provide more stringent constraints for astrophysics (e.g., Refs.~\cite{LIGOScientific:2018jsj, LIGOScientific:2020kqk, KAGRA:2021duu}),
cosmology (e.g., Refs.~\cite{LIGOScientific:2021aug, LIGOScientific:2023bwz}), and tests of general relativity (e.g., Refs.~\cite{LIGOScientific:2021sio, Magee:2023muf}).
Typically, a catalog of \ac{GW} observations are combined in a hierarchical manner \cite{Thrane:2018qnx, Mandel:2018mve, Vitale2020}, where the inferred properties of each source are compared to a model for their astrophysical population while accounting for selection biases.
For the population of binary \acp{BH} (BBHs) \acused{BBH} in particular, such hierarchical analyses have identified features and correlations in the distribution of source properties (see, e.g., Refs.~\cite{Callister:2021fpo, Biscoveanu:2022qac, Adamcewicz:2023mov, Godfrey:2023oxb, Heinzel:2023hlb, Heinzel:2024hva, Pierra:2024fbl, Li:2023yyt, Franciolini:2022iaa, Hussain:2024qzl, Antonini:2025zzw, Adamcewicz:2022hce, Sadiq:2025vly, Mould:2022ccw} and Ref.~\cite{Callister:2024cdx} for a review).

As \ac{GW} catalogs continue to grow it will become more computationally challenging to analyze them.
Current catalogs likely harbor non-astrophysical transients \cite{KAGRA:2021vkt};
if future catalogs continue to include low-significance triggers, unbiased population inference may eventually require simultaneously modeling the astrophysical and noise distributions \cite{Messenger:2012jy, Gaebel:2018poe, Galaudage:2019jdx, Ashton:2021tvz, Heinzel:2023vkq}.
Further, the computational cost of accurately estimating the population likelihood scales with the number of observed events \cite{Tiwari:2017ndi, Essick:2022ojx} and may be nearly quadratic in catalog size \cite{Talbot:2023pex}.
If the likelihood is misestimated, population inference may be biased.

Previously suggested methods to efficiently and accurately estimate the population likelihood include: fitting a density estimator to each set of single-event posterior samples \cite{Wysocki:2018mpo, Golomb:2021tll, Mancarella:2025uat, Bers:2025tei}; training conditional density estimators for the single-event posterior probability directly on \ac{GW} data \cite{Dax:2021tsq}; training density estimators conditioned on \ac{GW} catalogs \cite{Leyde:2023iof}; fitting an approximation to the selection function \cite{Talbot:2020oeu, Wong:2020wvd, Gerosa:2020pgy, Mould:2022ccw, Chapman-Bird:2022tvu, Callister:2024qyq, Lorenzo-Medina:2024opt}; or even performing inference on the detected distribution and reconstructing the astrophysical distribution of BBH in postprocessing \cite{Toubiana:2025syw}.
Fundamentally, however, each of these approaches introduces new approximations with systematic biases that may be difficult to quantify.

In this work, we evaluate a simple and complementary solution: modestly increasing the detection threshold and thus analyzing fewer events.
This both removes non-astrophysical transients from the catalog \textit{and} increases the accuracy of the population likelihood estimate.
In turn, it reduces the computational cost of \ac{GW} population analyses, potentially assisting analyses of future catalogs of thousands \cite{Kiendrebeogo:2023hzf} to hundreds of thousands \cite{Gupta:2023lga} of \ac{BBH} mergers.

However, if we are to propose analyzing fewer events, we must critically evaluate the balance between the astrophysical information we could gain from large catalogs of \ac{GW} observations and the associated growth in analysis complexity.
Ref.~\cite{Gaebel:2018poe} studied this balance in the context of joint analyses of non-astrophysical and astrophysical transients.
Using a simplified model for \ac{GW} detection and source parameter estimation, they found that a modest increase in the detection significance threshold may fortify inference of the \ac{BH} mass distribution against systematic biases from noise.
In the context of standard-siren cosmology, Ref.~\cite{Essick:2024uhl} used the Fisher matrix formalism \cite{Vallisneri:2007ev} to estimate how each observed binary merger informs our understanding of the redshift distribution of \ac{GW} sources.
For next-generation \ac{GW} observatories, Ref.~\cite{Essick:2024uhl} predicted that the loudest binary mergers will provide the majority of the information on the redshift distribution.
Ref.~\cite{Haster:2020sdh} also employed the Fisher matrix formalism, studying how constraints on the nuclear equation of state are informed by neutron stars merging at cosmological distances; there, they found that quieter sources from a redshift $z \sim 1$ are the most informative.
However, the Fisher information is only approximately correct for relatively loud signals \cite{Vallisneri:2007ev} and even then can over- or under-estimate uncertainties \cite{Zanolin:2009mk, Vitale:2010mr, Vitale:2011zx, Rodriguez:2013mla}.
Such inaccuracies propagate forward when forecasting measurements for population studies.

Here, we study how constraints on the astrophysical BBH population evolve with increasing detection thresholds, using a large catalog of simulated BBHs with full parameter estimation.
In Sec.~\ref{sec:methods}, we describe how we simulate mock catalogs of observed BBH mergers and infer the population parameters of their underlying astrophysical distribution.
Then, in Sec.~\ref{sec:stat-unc}, we describe how our measurement uncertainty on the population distribution evolves with the detection threshold.
In Sec.~\ref{sec:sys-unc}, we show that an increased significance threshold can result in reduced analysis complexity via more accurate estimation of the population likelihood.
Finally, in Sec.~\ref{sec:discussion}, we summarize our results and their implications for population analyses performed on future catalogs of BBH mergers.

\section{Methods} \label{sec:methods}

\subsection{\ac{GW} detection and characterization}

Given two time-series $a$ and $b$ of duration $T$ in a \ac{GW} interferometer indexed by $I$, their noise-weighted inner product is
\begin{align}
    \langle a, b \rangle = 4 \, \mathrm{Re} \left( \int_{f_{\min{}}}^{f_{\max{}}} \dd f \, \frac{\tilde{a}(f) \tilde{b}^*(f) }{ S^I(f) }  \right) \, ,
\end{align}
where $S^I(f)$ is the power spectral density characterizing the detector noise and $\tilde{a}$ and $\tilde{b}$ are the frequency-domain representations of $a$ and $b$.
The minimum and maximum frequencies, $f_{\min{}}$ and $f_{\max{}}$, are chosen based on the sensitive bandwidth of the detector and the Nyquist frequency of the data.

\ac{GW}s from BBH mergers have been detected with template-based matched filtering (e.g., Refs.~\cite{Messick:2016aqy, Sachdev:2019vvd, Hanna:2019ezx, 2021SoftX..1400680C, Adams:2015ulm, Aubin:2020goo, Allen:2004gu, Allen:2005fk, DalCanton:2014hxh, Usman:2015kfa, Nitz:2017svb} among others) outputting the measured \ac{SNR}.
Data $d^I$ are observed in each detector; assuming there is a \ac{GW} signal in those data described by a template \ac{BBH} signal $h^I$, the matched-filter \ac{SNR} measured in that detector is
\begin{align}
    \rho_{I} = \frac{\langle d^{I}, h^I(\theta) \rangle}{\sqrt{\langle h^I(\theta), h^I(\theta) \rangle}} \, .
\end{align}
Here, the 15 parameters $\theta$ include the BH masses, spin vectors, and relative orientation and location with respect to the detector.
For multiple detectors, the network \ac{SNR} is defined as the quadrature sum of the
\ac{SNR}s in each detector \cite{Jaranowski:1998qm, Finn:2000hj, Essick:2023toz, Gerosa:2024isl}
\begin{align} \label{eq:net-mf-snr}
    \rho^2 = \sum_{I} \rho_{I}^2.
\end{align}
Search pipelines also estimate how often noise transients produce triggers in the absence of astrophysical signals, thus producing a ranking of candidates according to the \ac{FAR}.

Confident triggers are defined as passing some predetermined threshold in the \ac{SNR} or \ac{FAR}.
Their source parameters are inferred using the Whittle likelihood for single events in Gaussian noise,
\begin{align} \label{eq:whittle}
    p(d^{I} | \theta) \propto \exp \left[ \langle d^{I}, h^I(\theta)\rangle - \frac{1}{2}\langle h^I(\theta), h^I(\theta) \rangle \right].
\end{align}
For coincident data in multiple \ac{GW} detectors, we multiply together the likelihoods of the data in each detector, yielding the single-event likelihood $p(d|\theta) = \prod_{I} p(d^{I} | \theta)$.
Using Bayes' theorem, we can compute the single-event posterior $p(\theta | d) \propto p(d|\theta) p(\theta)$ after adopting a prior $p(\theta)$.
For an overview of parameter estimation for compact binary mergers, see Ref.~\cite{Thrane:2018qnx}.

\subsection{Population inference with selection effects} \label{sec:methods-pop-inf}

With $\nobs{}$ detected BBH mergers, we can assemble a catalog of their observed data $\mathcal{D} = \{ d_i \}_{i=1}^{\nobs{}}$, each with a priori unknown parameters $\theta_i$.
The catalog is subject to a selection bias, as not all sources are detectable.
In the presence of selection effects, the likelihood of the observed catalog given an astrophysical population of merging BHs parametrized by $\Lambda$ can be written \cite{Thrane:2018qnx, Mandel:2018mve, Vitale2020}
\begin{align}
    p( \mathcal{D} | \Lambda) \propto \pdet{}(\Lambda)^{-\nobs{}} \prod_{i=1}^{\nobs{}} p(d_i | \Lambda) \, .
\end{align}
The likelihood of each observation is
\begin{align}
    p(d_i | \Lambda) = \int \dd{\theta_i}\,p(d_i | \theta_i) p(\theta_i | \Lambda) \, ,
    \label{eq:single-likelihood-given-lambda}
\end{align}
where the population model $p(\theta| \Lambda)$ is the astrophysical distribution of source properties as determined by population-level parameters $\Lambda$ that we aim to infer from the data.
The fraction $\pdet{}(\Lambda)$ of sources in the population
that are detectable is
\begin{align}
\pdet{}(\Lambda) = \int \dd{d} \dd{\theta}  \Theta( \rho(d) > \rho_* ) p(d |\theta) p(\theta | \Lambda)
\, ,
\end{align}
where $\rho(d)$ is a selection function of the data, $\rho_*$ is a threshold for selection, and $\Theta$ is the Heaviside function.
For real \ac{GW} catalogs, $\rho(d)$ may be the matched-filter \ac{SNR} between the observed data and best-matching signal template, the probability that an event is astrophysical in origin \cite{LIGOScientific:2018mvr, LIGOScientific:2021usb, Nitz:2021zwj, Mehta:2023zlk, Mehta:2025oge}\footnote{This probability is conditioned on assumptions about the distribution of astrophysical source properties; see Ref.~\cite{Banagiri:2023ztt} for an overview.} or the \ac{FAR} computed by search pipelines \cite{LIGOScientific:2018mvr, LIGOScientific:2021usb, KAGRA:2021vkt}.
In this work, we will take $\rho(d)$ to be the network matched-filter \ac{SNR} (Eq.~\eqref{eq:net-mf-snr}).

\subsection{Monte Carlo estimation and uncertainty}

Evaluating the detection function $\pdet{}$ for one choice of $\Lambda$ is extremely computationally expensive; doing so for many $\Lambda$ is intractable.
Instead, we estimate $\pdet{}$ first as a Monte Carlo integral over simulated sources $\catinj{} = \{ (d_m, \theta_m) \}_{m = 1}^{\ninj{}}$ drawn from one population, $\theta_m \sim p(\theta | \ldraw{})$, which are then importance sampled to $p(\theta | \Lambda)$ for all other $\Lambda$ \cite{Tiwari:2017ndi, 2019RNAAS...3...66F}.
That is,
\begin{align} \label{eq:monte-carlo-vt}
    \pdet{}(\Lambda) &\approx 
    \frac{1}{N_{\rm inj}} \sum_{k = 1}^{N_{\rm det}} \frac{p(\theta_k | \Lambda)}{p(\theta_k | \ldraw{})}
    \, , 
\end{align}
where \ninj{} is the total number of simulated sources and \ndet{} is the number that are detectable, i.e., those with $\rho(d_m) > \rho_*$.
The logarithm $\ln \pdet{} (\Lambda)$ of this Monte Carlo estimate has an associated variance
\begin{align}
    \dvar{} \approx \frac{1}{\neff} - \frac{1}{N_{\rm inj}} \, , \label{eq:vdet}
\end{align}
where \neff{} is the effective sample size in the Monte Carlo integral of Eq.~\eqref{eq:monte-carlo-vt} \cite{kish1995, Talbot:2023pex}.
Here, \dvar{} depends on $\Lambda$ and $\rho_*$ through \neff{}, and on \catinj{} through \neff{} and \ninj{}.

Similarly, the single-event likelihood is computationally expensive to evaluate.
Typically, the parameters of each event are inferred independently of other events, and so each confident detection has an associated set of \npe{} posterior samples $\{ \theta_{ij} \}_{j = 1}^{N_{\rm PE}} \sim p(\theta_i | d_i)$.
We then estimate Eq.~\eqref{eq:single-likelihood-given-lambda} as a Monte Carlo integral over $\theta_{ij}$, after replacing $p(d_i | \theta_i)$ with $p(\theta_i | d_i) / p(\theta_i)$ via Bayes' theorem (up to a normalization constant) and importance sampling from the single-event prior to the population prior.
Our estimate of Eq.~\eqref{eq:single-likelihood-given-lambda} has a Monte Carlo variance \pevar{} with a similar form as Eq.~\eqref{eq:vdet}, where \pevar{} is inversely proportional to the effective sample size in the Monte Carlo integral of Eq.~\eqref{eq:single-likelihood-given-lambda} and \npe{}.

In total, our estimate of $\ln p(\mathcal{D}|\Lambda)$ has a variance
\cite{Essick:2022ojx, Talbot:2023pex}
\begin{align} \label{eq:like-var}
    \mathcal{V} \approx \sum_{i=1}^{\nobs{}} \pevar{} + \nobs^2 \dvar{} = \vpe{} + \vdet{} \, ,
\end{align}
where we define $\vpe{} = \sum_{i = 1}^{\nobs{}} \pevar{}$ and $\vdet{} = \nobs^2 \dvar{}$.
When this variance is large, the estimated population likelihood may assign high probability to regions of the parameter space that are inconsistent with the astrophysical distribution implied by the observed catalog.
We can reduce \lvar{} by increasing \ninj{} and \npe{}, although this is not always feasible due to the computational cost of simulating many \ac{GW} signals when estimating \pdet{} or drawing samples from each single-event posterior.
In practice, we can regularize the likelihood during sampling or post-processing by excluding all $\Lambda$ that yield a variance larger than some threshold $V$ \cite{Talbot:2023pex, jack_math_tome}.
However, this is a non-trivial modification of the likelihood and should only be done if necessary, e.g., when the variance correlates strongly with $\Lambda$ within the posterior support.

Here, our goal is to determine whether raising the detection threshold is a reasonable strategy for regularizing Monte Carlo estimates of the likelihood.
Higher \ac{SNR} thresholds naturally reduce \vpe{} by reducing \nobs{}.
Generically, \vdet{} does not decrease with \ac{SNR} threshold.
In App.~\ref{app:nmin-scaling},
we derive a heuristic scaling relationship with SNR threshold, $\vdet{} \propto \rho_*^{b - 2a}$ if $\nobs{} \propto \rho_*^{-a}$ and $\ndet{} \propto \rho_*^{-b}$ for some powers $a, b$ (with $a = b = 3$ in a flat universe where the merger rate is constant in luminosity distance and uncorrelated with other source properties) \cite{Schutz:2011tw, Chen:2014yla};
thus, \vdet{} declines with \ac{SNR} threshold for $a > b / 2$.
We anticipate $a \approx b \approx 3$ for real observing scenarios where the distribution of simulated sources---while broad so as to support many possible astrophysical distributions---is informed by previous catalogs \cite{o4a-injection-campaign}.

\subsection{Mock catalogs} \label{sec:methods-mock}

To construct a mock catalog of simulated \ac{GW} observations of merging BBHs, we jointly draw sources with parameters $\theta \sim p(\theta | \ltrue{})$ and realizations of noise in each detector.
For our astrophysical population $p(\theta | \ltrue{})$, we adopt phenomenological models from Ref.~\cite{KAGRA:2021duu}:
the distribution of primary BH masses $m_1$ is a tapered power-law continuum with a secondary Gaussian peak and the mass ratios $q$ follow a tapered power law \cite{Talbot:2018cva};
the merger rate evolves as a power law in $1+z$ over redshift $z$ \cite{Fishbach:2018edt};
the dimensionless component spin magnitudes $a_{1, 2}$ are independently and identically distributed from nonsingular Beta distributions \cite{Wysocki:2018mpo}; and the component spin tilt angles (relative to the orbital angular momentum) $\theta_{1,2}$ follow a mixture between uniform and Gaussian components in $\cos \theta_{1,2}$
with a peak at $\cos \theta_{1,2} = 1$ \cite{Vitale:2015tea, Talbot:2017yur}.
The exact population parameters $\ltrue{}$ that we adopt are the maximum-likelihood values from Ref.~\cite{KAGRA:2021duu} and can be found in App.~\ref{app:pop-pars}.
For each source with parameters $\theta$, the \ac{GW} signal $h(\theta)$ is generated with the \textsc{IMRPhenomXP} waveform approximant \cite{Pratten:2020ceb}.
Finally, these signals are added to data simulated in the LIGO Hanford, LIGO Livingston, and Virgo detectors, for which we adopt power spectral densities corresponding to their \ac{O4} design sensitivities \cite{T2000012}.

In this work, we take the selection function $f(d)$ to be the network matched-filter \ac{SNR} $\rho$ from Eq.~\eqref{eq:net-mf-snr}, evaluated between the data containing the simulated signal and the simulated signal itself.
This is no longer a purely data-dependent selection criterion \cite{Essick:2023upv} as it depends also on the true source properties, although we expect the induced bias to be minimal even for catalogs of $\mathcal{O}(10^3)$ events \cite{Mould:2025dts, Vitale:2025lms}.
Events are included in our mock catalog if $\rho > 11$; this threshold is calibrated \cite{Essick:2023toz} to roughly match a \ac{FAR} threshold of $1~\rm yr^{-1}$ used for BBH population analyses of the third gravitational-wave transient catalog (GWTC-3) \cite{KAGRA:2021duu}\footnote{This calibration is not exact for our study, as Ref.~\cite{Essick:2023toz} calibrated the matched-filter \ac{SNR} to \ac{FAR} assuming detector sensitivities from the \ac{LVK}'s third observing run, not the simulated \ac{O4} sensitivities we use.}.\acused{GWTC-3}\acused{GWTC-4}
In total, we select 1600 simulated events that satisfy $\rho > 11$ to make up our mock catalog.
For each of these, we infer the posterior distribution of source parameters assuming the same waveform model, detector network, and detector sensitivities as when simulating the strain data.
We sample the single-event posteriors using \texttt{dynesty} \cite{Speagle:2019ivv} and \texttt{bilby} \cite{bilby_paper}.
We use 4000 live points, yielding $\gtrsim$~15000 posterior samples for each source, allowing us to later reduce the Monte Carlo variance \pevar{} from each mock event.
For priors and waveform settings, see App.~\ref{app:single-event-settings}.

To estimate $\pdet{}(\Lambda)$, we generate and select additional simulated data realizations in the same manner as we constructed our mock catalog.
However, we instead draw source properties from a distribution $p(\theta | \ldraw{})$, which takes the same form as our underlying astrophysical population in $m_1$, $q$, and $z$, flat distributions in $a_{1,2}$ and $\cos \theta_{1,2}$, and $\ldraw{}$ chosen to ensure $\mathrm{supp}~p(\theta | \Lambda) \subseteq \mathrm{supp}~p(\theta | \ldraw{})$ for all $\Lambda$ of interest; see App.~\ref{app:pop-pars} for the exact values of $\ldraw{}$.
For computational efficiency, we also interpolate the most optimistic estimate of the \ac{SNR} as a function of the source-frame total mass and redshift, assuming that no noise is present in the data and that the \ac{BBH} is otherwise configured to maximize the observed power in the detector network.
In particular, we compute the \ac{SNR} in each detector $\rho_I$ assuming that the \ac{BBH} is directly overhead and oriented face-on to that detector, and then add those in quadrature according to Eq.~\eqref{eq:net-mf-snr}.
We also assume the \acp{BH} have equal masses, and maximal and orbit-aligned spins.
We immediately reject sources for which the most-optimistic \ac{SNR} is $< 11$, as it is highly unlikely that their matched-filter \ac{SNR} will be scattered by noise above 11.
For additional details, see Ref.~\cite{Vitale:2025lms}.
In total, we simulate $N_{\rm inj} \approx 6.4 \times 10^8$ events, of which $N_{\rm det} = 1.1 \times 10^7$ satisfy $\rho > 11$.

We sample the population posterior $p(\Lambda|\mathcal{D}) \propto p(\mathcal{D}|\Lambda) p(\Lambda)$ with \texttt{dynesty} and \texttt{bilby}, using the population likelihood $p(\mathcal{D}|\Lambda)$ as implemented in \texttt{gwpopulation} \cite{2019PhRvD.100d3030T, Talbot:2024yqw} with \texttt{JAX} \cite{jax2018github} to run on GPUs.
We repeat population inference for catalogs selected under a series of increasingly stringent thresholds, $\rho_* \in \{ 11, 12, 13, ..., 40 \}$.
These catalogs are generated by down-selecting events from our initial catalog of 1600 sources; we similarly threshold the simulated sources used to estimate selection effects.
In all analyses, we take uniform priors $p(\Lambda)$ that are broad enough to encompass the posterior support for all population analyses in this work; see App.~\ref{app:pop-pars}.
We prevent large uncertainties in the Monte Carlo approximations of the population likelihood by imposing a limit on variance $\lvar{} < V = 4$ during inference.
In practice, \lvar{} consistently lies below this limit over the posterior support in all of our analyses (see Fig.~\ref{fig:joint-var-marginals}, although there may be spurious modes in the likelihood with $\lvar > 4$ that we have excised).

Our mock catalog of parameter estimation, simulations for sensitivity estimates, and population inference results can be found at Ref.~\cite{faroutdata}\footnote{\href{https://doi.org/10.5281/zenodo.17080422}{https://doi.org/10.5281/zenodo.17080422}}.

\section{Statistical Uncertainty} \label{sec:stat-unc}

Some population parameters may have an explicit astrophysical meaning, like the location of the Gaussian peak in primary masses, which was originally introduced to capture a predicted pileup of \ac{BH} masses caused by pulsational pair-instability supernovae \cite{Talbot:2018cva} (although this interpretation may now be disfavored \cite{Golomb:2023vxm, Hendriks:2023yrw, Roy:2025ktr}).
Typically, however, it is the overall inferred shape of the population distribution that can be compared to theoretical predictions of the \ac{BBH} populations produced via particular formation channels (e.g., Refs.~\cite{Broekgaarden:2021efa, vanSon:2022myr, Bavera:2022mef, Ye:2025ano}, among others).
Therefore, we inspect the \acp{PPD}, as opposed to posteriors on the population parameters, which we record in App.~\ref{app:hyperpars}.
The \ac{PPD} is the set of $p(\theta | \Lambda)$ with $\Lambda \sim p(\Lambda | \mathcal{D})$.
We quantify the growth in measurement uncertainty as $\rho_*$ increases via the width of the 90\% equal-tailed \ac{CI} of the \ac{PPD} relative to the width of the same interval of the \ac{PPD} when analyzing all 1600 events.
This ratio, which we denote \relunc{}, only accounts for the relative uncertainty of the posterior distributions, but not their relative locations.
However, the marginal \acp{PPD} for each analysis are typically centered on similar densities, allowing us to meaningfully compare \relunc{} between different analyses at particular locations in the source parameter space.

For clarity, we highlight only a subset of our analyses performed on catalogs with \ac{SNR} thresholds $\rho_* \in \{ 11, 12, 13, 14, 15, 20, 30\}$ (results for all analyses with $\rho_* \in \{ 11, \ldots 40 \}$ are included in the data release at Ref.~\cite{faroutdata}).
These thresholds are chosen to span a range of catalog sizes, from our largest catalog of 1600 events to a catalog of only 54 events comparable in size to \ac{GWTC-3} \cite{KAGRA:2021vkt}.
Catalog sizes for selected analyses are listed in Tab.~\ref{tab:nobs}; we also show the number of detectable simulations at each threshold.
For each analysis, we present results for redshift, primary mass, mass ratios, spin magnitudes, and spin tilts.

Finally, in App.~\ref{app:nobs69-additional} we repeat population inference with a fixed catalog size at a few \ac{SNR} thresholds to disentangle whether catalog size or \ac{SNR} drives constraints on the \ac{BBH} distribution.

\begin{table}
    \centering
    \begin{tabular}{ | c || *{7}{>{\centering\arraybackslash}p{0.7cm}|} }
        \hline
        $\rho_*$ & 11 & 12 & 13 & 14 & 15 & 20 & 30 \\
        \hline
        $\nobs{}$ & 1600 & 1205 & 904 & 705 & 560 & 221 & 54 \\
        \hline
        $\ndet{} [\times 10^7]$ & 1.1 & 0.81 & 0.62 & 0.48 & 0.38 & 0.15 & 0.041 \\
        \hline
    \end{tabular}
    \caption{
    Number of events $\nobs{}$ in our mock catalog, selected as a function of the matched-filter \ac{SNR} threshold $\rho_*$.
    Note $\nobs{} \propto \rho_*^{-3.21}$ and $\ndet{} \propto \rho_*^{-3.25}$.
    }
    \label{tab:nobs}
\end{table}

\subsection{Redshift} \label{sec:redshift}

\begin{figure}
    \centering
    \includegraphics[width=0.99\linewidth]{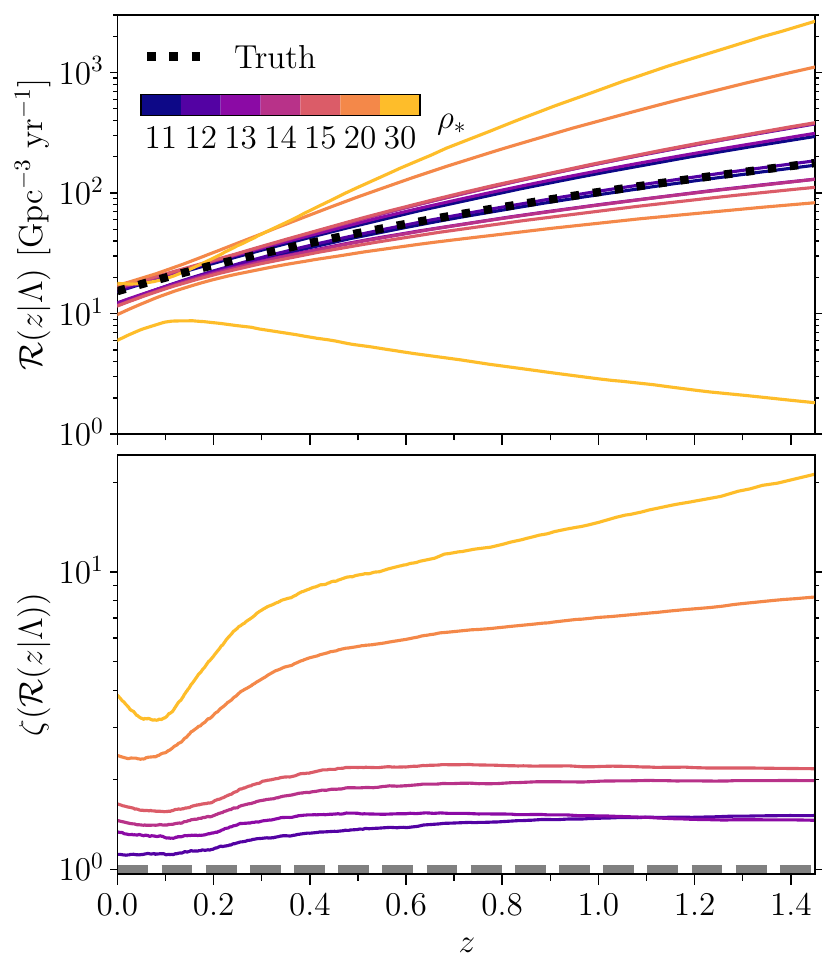}
    \caption{PPDs of the merger rate density $\mathcal{R}$ over redshift $z$ (top) inferred from catalogs with increasing detection threshold on the \ac{SNR}, $\rho_*$, indicated by colors from purple to yellow.
    Colored lines enclose the 90\% \ac{CI} and the dotted black line indicates the true distribution.
    In the bottom panel, we plot the width \relunc{} of the 90\% \ac{CI} inferred in each analysis relative to our analysis at $\rho_* = 11$.
    We include a gray dashed line at $\relunc{} = 1$ (i.e., no growth in statistical uncertainty with increased $\rho_*$) for reference.
    }
    \label{fig:mock-ppds-z}
\end{figure}

In our analyses, we model the astrophysical merger rate density $\mathcal{R}$ as a power law $\mathcal{R}(z) = \mathcal{R}_0 (1+z)^{\lambda_z}$ over redshift $z$ with index $\lambda_z$.
The true merger rate $\mathcal{R}_0$ at $z = 0$ is determined by the size of our mock catalogs, the sensitivity of our simulated detectors, and our assumed observing time.
As in Ref.~\cite{Mould:2025dts}, we assume an observing period of ten years, such that $\mathcal{R}_0 \approx 15.4\,\rm Gpc^{-3}\,yr^{-1}$, consistent with current constraints on the merger rate from the \ac{LVK}~\cite{KAGRA:2021duu}.

The 90\% \ac{CI} for the posterior of merger rate over redshift are shown in the top panel of Fig.~\ref{fig:mock-ppds-z} for select \ac{SNR} thresholds $\rho_*$.
In all analyses, we correctly recover the true evolution of the merger rate over redshift within at least 90\% credibility.
Further, even as the \ac{SNR} threshold is raised to $\rho_* = 20$, satisfied by only $13\%$ of the sources in our original mock catalog, the rate distributions allowed at 90\% credibility remain qualitatively consistent.
However, by $\rho_* = 30$, the uncertainty in rate is sufficiently large as to alter the conclusions we would draw from these results, in particular that the merger rate increases with increasing redshift.

Beyond qualitative differences in the \acp{PPD} for each analysis, we compare the statistical uncertainty achieved under each \ac{SNR} threshold to our analysis of all 1600 events in the bottom panel of Fig.~\ref{fig:mock-ppds-z}.
For all analyses, \relunc{} is smallest at $z \sim 0.1$, where our constraints on $\mathcal{R}$ are tightest.
At higher redshifts, \relunc{} grows with increasing $z$ and it grows more quickly with $z$ as $\rho_*$ increases.
When $\rho_*$ is raised from 11 to 13 (and \nobs{} is correspondingly nearly halved), \relunc{} increases by at most a factor of $\sim 1.5$ across all redshifts.
Increasing $\rho_*$ further to 15 doubles statistical uncertainty beyond $z \sim 0.1$.
For catalogs selected according to the highest \ac{SNR} thresholds shown in Fig.~\ref{fig:mock-ppds-z}, $\rho_* = 20$ and 30 ($\nobs = 221$ and 54), we find $\relunc \sim 8$ and $\relunc \sim 20$ at $z \sim 1.5$, respectively.
This is because a higher \ac{SNR} threshold tends to remove sources merging at higher $z$ that would otherwise help to constrain the evolution of the merger rate; see App.~\ref{app:nobs69-additional} for further discussion.

\subsection{Masses} \label{sec:primary-mass}

\begin{figure*}
    \centering
    \includegraphics[width=0.99\linewidth]{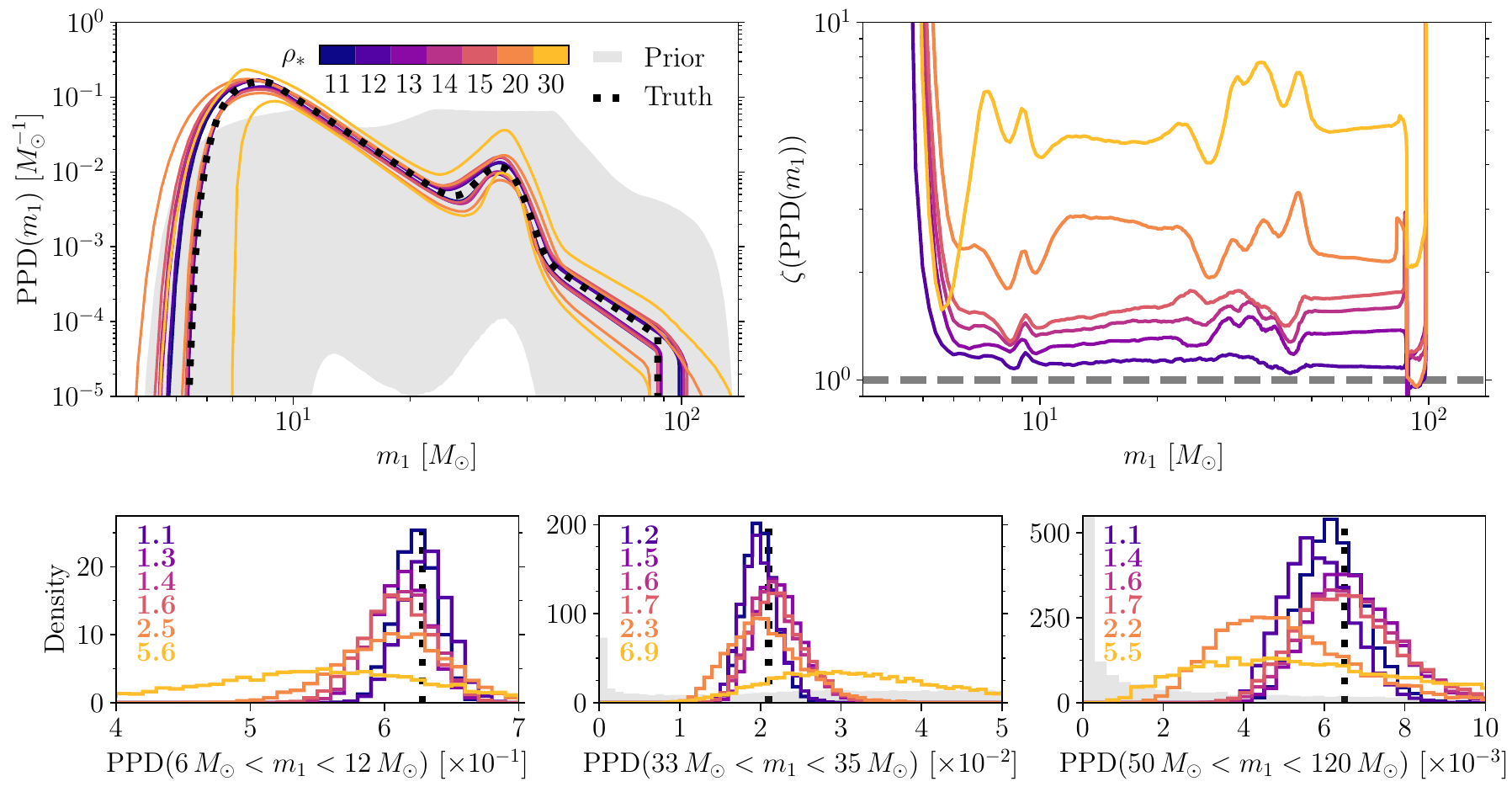}
    \caption{
    \acp{PPD} of the primary mass $m_1$ (top left) inferred from catalogs with increasing detection threshold on the SNR $\rho_*$.
    Colors and lines have the same meaning as Fig.~\ref{fig:mock-ppds-z}.
    Additionally, the gray region fills in the 90\% \ac{CI} of the prior.
    In the top right, we show the width of the 90\% \ac{CI} of the \acp{PPD} relative to our analysis at $\rho_* = 11$.
    In the bottom row, we show the posterior uncertainties on the integrated posterior population distributions over three subdomains of interest in $m_1$.
    Numbers on the left of each panel, colored according to \ac{SNR} threshold, note the 90\% \ac{CI} width, relative to our analysis at $\rho_* = 11$, of the posterior on the fraction of events in each subdomain.
    Note, for $6\,M_\odot< m_1 < 12\,M_\odot$, the \acp{PPD} mostly lie above the prior in density, and so no prior is visible in the bottom left panel.
    Conversely, the prior on the fraction of events lying in $50\,M_\odot< m_1 < 120\,M_\odot$ is strongly peaked at zero due to the sharp cutoff in our assumed population model which sends the prior population density to zero at high masses.
    }
    \label{fig:mock-ppds-m1}
\end{figure*}

In the top-left panel of Fig.~\ref{fig:mock-ppds-m1}, we show the \acp{PPD} and prior population distributions for $m_1$ for select \ac{SNR} thresholds $\rho_*$.
All of our analyses successfully recover the true distribution.
Generally, the inferred population distributions are comparable up to $\rho_* = 15$, with similar measurements and uncertainties on the extrema of the $m_1$ distribution, the shape of the power-law continuum, and the location and width of the peaks at $m_1 \lesssim 10\,M_\odot$ and $m_1 \approx34\,M_\odot$.
Meanwhile, by $\rho_* = 30$ the measurement uncertainty on all of these features is noticeably larger.

In the top right panel of Fig.~\ref{fig:mock-ppds-m1}, we show the measurement uncertainty achieved in each analysis relative to our analysis of all 1600 events satisfying $\rho_* = 11$.
For $m_1 \lesssim 7\,M_\odot$, we observe that \relunc{} is larger when $\rho_* = 20$ than when $\rho_* = 30$.
This is because the location of the \ac{PPD} at low primary mass and high \ac{SNR} threshold is more susceptible to Poisson fluctuations, as fewer low mass sources are loud enough to be included in the catalog.
For $m_1 \gtrsim 87\,M_\odot$, the sharp cutoff at high masses in our population model causes \relunc{} to dip and then diverge as the lower and upper bounds of the $90\%$ \ac{CI} of the probability density in our analysis with $\rho_* = 11$ drop to zero.

For $7\,M_\odot \lesssim m_1 \lesssim 87\,M_\odot$, \relunc{} is only $\sim 20\%$ larger when the \ac{SNR} threshold is raised from 11 to 12 or 13;
when $\rho_*$ is raised to 15, the 90\% \ac{CI} width typically grows by $\sim 60\%$, and at most $80\%$ (at $m_1 \approx 32\,M_\odot$).
This is despite only one third of our sources satisfying $\rho > 15$.
Measurement uncertainties double when the \ac{SNR} threshold is raised to $\rho_* = 20$ ($\nobs = 221$); a more dramatic increase if the threshold to $\rho_* = 30$ reduces our catalog to only 54 events and raises the relative uncertainty by a factor of $\sim$5--7.

We can also ask more specific questions of the population distribution.
For example, Ref.~\cite{Vitale:2025lms} identified that it may be possible to measure the fraction of sources in particular domains even if the broad shape of the population remains somewhat unconstrained.
Here, we compute the posterior on the fraction of sources with primary masses in $[a, b]$ by integrating the \acp{PPD} over those domains, which we denote $\mathrm{PPD}(a < m_1 < b)$.
In the bottom row of Fig.~\ref{fig:mock-ppds-m1}, we show this between 6--$12\,M_\odot$, 33--$35\,M_\odot$, and 50--$120\,M_\odot$.
We observe a consistent pattern in all three plots.
Raising the \ac{SNR} threshold from $\rho_* = 11$ to $15$, our uncertainty (quantified by the 90\% \ac{CI}) on the fraction of events lying in each domain of $m_1$ grows by no more than a factor of 1.7.
Increasing our threshold to $\rho_* = 20$ roughly doubles our measurement uncertainty; by $\rho_* = 30$, our uncertainties are a factor of 5.5--6.9 times as large.

\begin{figure}
    \centering
    \includegraphics[width=0.99\linewidth]{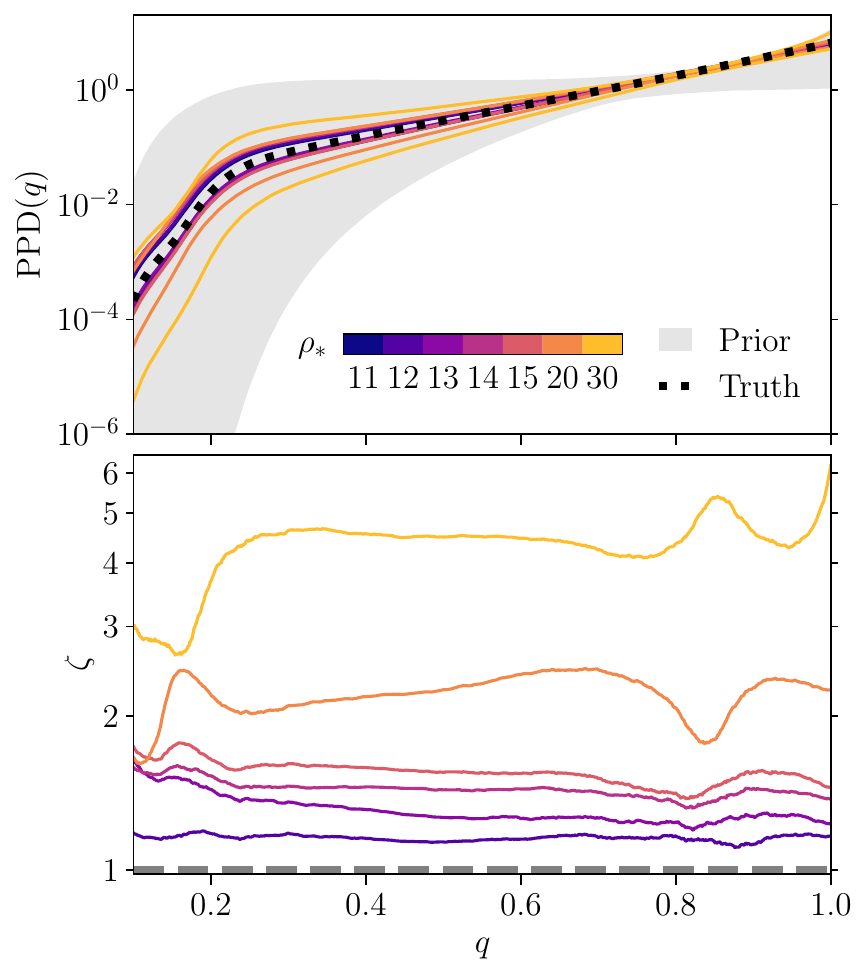}
    \caption{
    The 90\% \ac{CI} of the \acp{PPD} in mass ratio $q$ (top) and width of the 90\% \ac{CI} of the \acp{PPD} relative to our analysis at $\rho_* = 11$ (bottom) at select thresholds $\rho_*$.
    Thresholds, shading, lines, and colors match those in the top row of Fig.~\ref{fig:mock-ppds-m1}.
    }
    \label{fig:mock-ppds-q}
\end{figure}

We show \acp{PPD} for the mass ratio $q$ in the top panel of Fig.~\ref{fig:mock-ppds-q} at select \ac{SNR} thresholds $\rho_*$.
We recover the true population distribution in $q$ within at least the 90\% \ac{CI} in all analyses.
In the bottom panel of Fig.~\ref{fig:mock-ppds-q}, we compare the width of the 90\% \ac{CI} at each $\rho_*$ to that achieved in our analysis of all 1600 events with $\rho_* = 11$.
When $\rho_*$ is raised from 11 to 15, \relunc{} is at most $\sim 1.8$ (at $q \sim 0.1$--0.2).
When $\rho_*$ is raised from 11 to 20, the statistical uncertainty roughly doubles (with the largest \relunc{} of $\sim 2.5$ at $q \sim 0.7$).
When the \ac{SNR} threshold is raised further to $\rho_* = 30$, \relunc{} is as large as $\sim 6.2$ (at $q = 1$).
That the statistical uncertainty on the mass-ratio distribution only increases by $\mathcal{O}(10\%)$ when the \ac{SNR} threshold is raised from $\rho_* = 11$ to 15 mirrors the same trend in our constraints on the primary-mass distribution.
This is in part because the mass-ratio distribution is conditioned on the primary mass---so a well-constrained primary mass distribution lends itself to tight constraints on the mass-ratio distribution.

\subsection{Spin magnitudes and tilts}
\label{sec: spins}

\begin{figure*}
    \centering
    \includegraphics[width=0.99\linewidth]{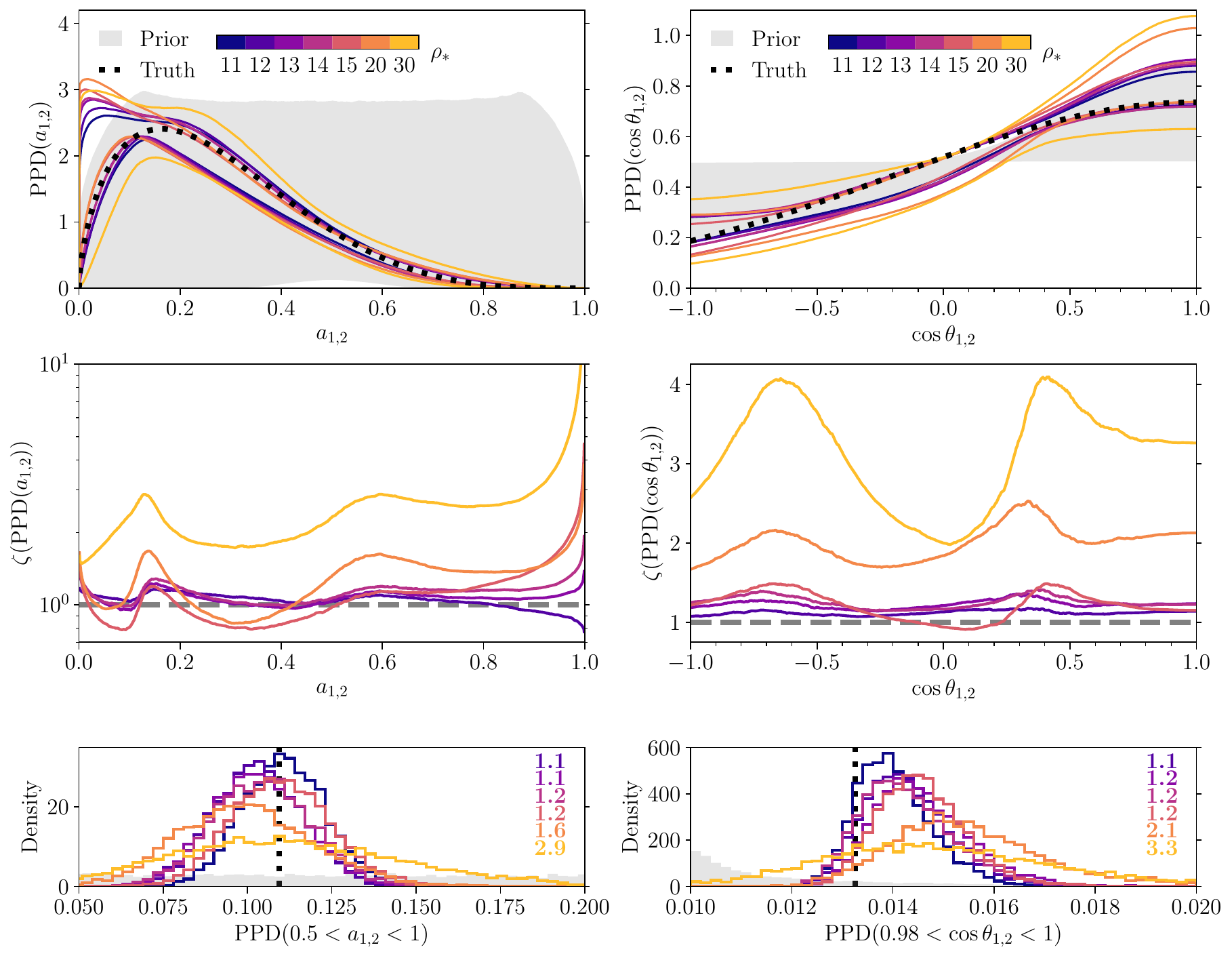}
    \caption{
    \acp{PPD} (top row), width of the 90\% \ac{CI} of the \acp{PPD} relative to our analysis at $\rho_* = 11$ (middle row), and \acp{PPD} integrated over select domains (bottom row) for spin magnitudes $a_{1,2}$ (left column) and tilts $\cos \theta_{1,2}$ (right column). Thresholds, shading, lines, and colors match those in Fig.~\ref{fig:mock-ppds-m1}.
    }
    \label{fig:mock-ppds-spins}
\end{figure*}

In the top-left panel of Fig.~\ref{fig:mock-ppds-spins}, we show \acp{PPD} for the identically-distributed component spin magnitudes, $a_1$ and $a_2$.
Although the truth sometimes lies outside the 90\% \ac{CI} for the analyses shown, they all recover the truth within the $\sim99\%$ \ac{CI}.
Among the analyses shown in Fig.~\ref{fig:mock-ppds-spins}, our constraints on the spin magnitude population distribution have similar locations and shapes, with a slight increase in the width of the credible region as we increase $\rho_*$ to 20 and 30.

In the middle-left panel of Fig.~\ref{fig:mock-ppds-spins}, we quantify the relative constraining power of each analysis via \relunc{}.
Here, we observe a striking result: our analyses up to $\rho_* = 20$ yield essentially the same measurement uncertainty relative to our analysis at $\rho_* = 11$.
At $\rho_* = 30$, \relunc{} is at most $\sim$2--3 (at both $a_{1,2} \sim 0.15$ and $a_{1, 2} \sim 0.6$). 
We note that, although raising the SNR threshold reduces the number of analyzed events, \relunc{} sometimes falls below one. This apparent increase in certainty results from our choice of population model.
Specifically, the nonsingular beta distribution shifts density from $a_{1,2} > 0$ to increase it near $a_{1,2} \sim 0$. Since our analyses with $\rho_* > 11$ prefer a spin distribution with more support near zero, the 90\ CI of the PPD shrinks at higher spin magnitudes.

Of particular astrophysical relevance is the proportion of events with large spins, which may be indicative of, e.g., hierarchical mergers \cite{Gerosa:2021mno}.
For the purpose of demonstration, we compute the posterior on the fraction of sources with $a_{1,2} > 0.5$, shown in the bottom-left panel of Fig.~\ref{fig:mock-ppds-spins}.
There, we see that we correctly recover the true fraction with all analyses.
Further, other than slight scatter about the truth, the inferred fractions of sources with spins $> 0.5$ is consistent within measurement error up to $\rho_* = 15$.
Our uncertainty grows slightly when $\rho_*$ is increased to 20 and moreso by $\rho_* = 30$, with 90\% \acp{CI} a factor of 1.6 and 2.9 larger than at $\rho_* = 11$, respectively.

Overall, we observe that reducing our catalog size from 1600 to 560 or even 221 events---by increasing $\rho_*$ to 15 or 20 (cf. Tab.~\ref{tab:nobs})---does not meaningfully change the conclusions we draw about the distribution of spin magnitudes.

We note that the agreement in population-level measurement uncertainties at high spins may be driven in part by the population model itself.
In order to match the relatively high density of sources with small spins implied by our mock catalogs, the non-singular beta distribution must remove density from the region of high spins.
Using a flexible model for the spin-magnitude distribution (e.g., Refs.~\cite{Golomb:2022bon, Edelman:2022ydv, Godfrey:2023oxb}) would likely increase the uncertainty on the shape, as is the case for all of our models (see similar comparisons of truncated Gaussians versus flexible models for the spin distribution in the \ac{LVK} population analysis of \ac{GWTC-4} \cite{LIGOScientific:2025pvj}).

We show \acp{PPD} on the component spin tilts in the top-right panel of Fig.~\ref{fig:mock-ppds-spins}.
Here, we observe that all of our analyses correctly recover the true distribution; further, the location and width of the 90\% credible region is essentially the same up to $\rho_* = 15$.
We show the uncertainty relative to our analysis of all 1600 sources in the middle-right panel of Fig.~\ref{fig:mock-ppds-spins}; we find that \relunc{} only doubles at most spin tilts when the \ac{SNR} threshold is raised to $\rho_* = 20$.
Following Ref.~\cite{Vitale:2025lms}, we compute posteriors on the fraction of sources with tilt angles $\lesssim 10^\circ$ as small (large) tilt angles may be indicative of field (dynamical) formation environments \cite{Rodriguez:2016vmx, Marchant:2021hiv}\footnote{Note, however, that \ac{BBH} formed in the field could be born with large tilt angles depending on the details of supernova kicks and fallback (e.g., Refs.~\cite{Steinle:2020xej, Baibhav:2024rkn}) as well as three-body interactions (e.g., Refs.~\cite{Liu:2017yhr, Yu:2020iqj}).}.
We show posteriors on the fraction of sources with $\cos \theta_{1,2} > 0.98$ (corresponding to $\theta_{1,2} \lesssim 10^\circ$) in the bottom-right panel of Fig.~\ref{fig:mock-ppds-spins}.
Here, we recover the true fraction in all of our analyses.
Since the \acp{PPD} (top-right) agree well up to $\rho_* = 15$ near $\cos \theta_{1,2} = 1$, we also observe that the posterior on the fraction of sources with $\cos \theta_{1,2} > 0.98$ is essentially the same up to $\rho_* = 15$; meanwhile, the posteriors at $\rho_* = 20$ and $\rho_* = 30$ are somewhat broader, with 90\% \acp{CI} that are a factor of 2.1 and 3.3 larger than at $\rho_* = 11$, respectively.
We note that \relunc{} falls below one at $\rho_* = 15$ for $\cos \theta_{1,2} \sim 0.1$.
This is because the analysis at $\rho_* = 15$ yields slightly tighter constraints on the fraction $\xi_{\cos \theta}$ of sources in the peak at $\cos \theta_{1,2} = 1$ than analyses at lower \ac{SNR} threshold, with less support at small values of $\xi_{\cos \theta}$ (cf. Fig.~\ref{fig:mock-boxy-spin}).
In turn, tighter constraints on $\xi_{\cos \theta}$ imply a more confident measurement of the tilt distribution around the peak, $\cos \theta_{1,2} \gtrsim 0$.
That $\xi_{\cos \theta}$ is slightly better measured at $\rho_* = 15$ than at lower thresholds (with larger $\nobs$) is possibly due to randomness in catalog membership as the \ac{SNR} threshold is raised.

Like primary mass and spin magnitudes, the astrophysical conclusions we might draw from the distribution of cosine-spin tilts are unchanged whether we analyze all 1600 events satisfying $\rho_* = 11$, or only 560 events satisfying $\rho_* = 15$.
However, we must caution that---like with spin magnitudes---our model for the spin tilt distribution is relatively strong.
Refs.~\cite{Vitale:2022dpa, Vitale:2025lms} relax the assumption that the location parameter of the Gaussian component is pinned at $\cos \theta_{1,2} = 1$ and recover the location of the Gaussian with large uncertainty, even for mock catalogs of up to 1500 events.
Therefore, if we repeated our study with a more flexible population model for $\cos \theta_{1,2}$, we expect our statistical uncertainty on the structure of the cosine-tilt distribution to grow more quickly with increasing threshold and decreasing catalog size than found here.

Our results may also indicate that spin population inference is most informed by high \ac{SNR} sources; see App.~\ref{app:nobs69-additional} for additional discussion.

\section{Systematic Uncertainty} \label{sec:sys-unc}

For computational efficiency in our analyses, we estimated the population likelihood via Monte Carlo integration, which has an associated statistical uncertainty quantified by the variance of the estimator, \lvar{}; see Eq.~\eqref{eq:like-var}.
Population inference may be biased if \lvar{} is large, as the estimated likelihood may be far from the true likelihood.
This variance can be reduced by increasing the number of samples used for Monte Carlo integration;
see Eq.~\eqref{eq:vdet} and Fig.~5 in Ref.~\cite{Talbot:2023pex}.
However, while tractable for our largest catalog size of 1600 simulated sources, increasing \npe{} or \ninj{} could quickly become \textit{intractable} for catalogs of thousands or tens of thousands of events, as expected from next-generation \ac{GW} detectors \cite{Gupta:2023lga}.

The variance of our estimate of the population likelihood can also be reduced by decreasing catalog size, which we accomplish in this work by increasing the \ac{SNR} threshold for inclusion in the catalog.
In this section, we first evaluate the tradeoff between the measurement uncertainty of population inference and the Monte Carlo uncertainty in the estimate of the likelihood as we increase the \ac{SNR} threshold.
The selection variance dominates \lvar{} for the analyses we considered (see App.~\ref{app:npe}), so we then compute the minimum number of simulations required to accurately estimate the detection efficiency $\pdet{}$ as a function of \ac{SNR} threshold, alongside an estimate for the associated computational cost.

\subsection{Trading systematic for statistical uncertainty} \label{sec:tradeoff}

\begin{figure*}
    \centering
    \includegraphics[width=0.99\linewidth]{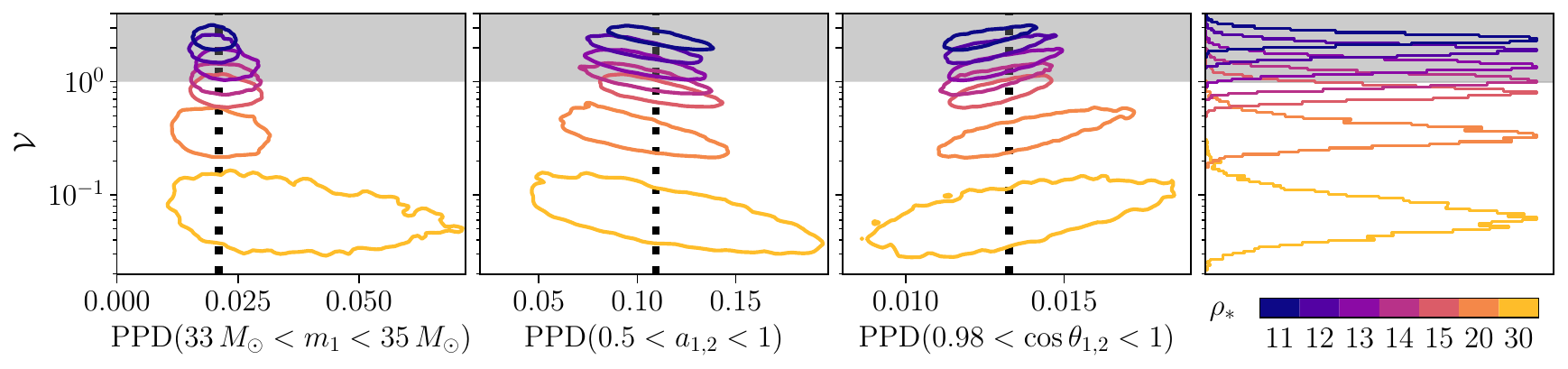}
    \caption{
    Joint posterior distributions for the total variance of the log-likelihood estimator \lvar{} and integrated \acp{PPD}.
    In the right panel, we histogram (with arbitrary normalization) the marginal posterior on the log-likelihood estimator variance.
    In all other panels, we show the 90\% credible regions of joint distributions on \lvar{} and the fraction of sources lying in select domains of primary mass, spin magnitude, or cosine-spin tilt.
    Colors from cooler to warmer denote the threshold $\rho_*$ and dotted black lines denote values in the true population.
    }
    \label{fig:joint-var-marginals}
\end{figure*}

Particularly for \ac{BBH} masses and spins, we have found that raising the selection threshold as high as $\rho_* \sim 15$ does not dramatically change the uncertainty on the measured population distribution.
Simultaneously, uncertainty in the likelihood estimator is lower for smaller catalogs; further, higher detection thresholds will typically exclude more noise transients, increasing the purity of the catalog.

We note that \lvar{} itself does not quantify a systematic uncertainty in a population analysis, although it does induce systematic uncertainty.
Precisely how uncertain our estimate of the population likelihood can become before biasing our results depends in detail on the true astrophysical population and the population model used to fit the data.
Ref.~\cite{Talbot:2023pex} originally noted that inference may be noticeably biased when the variance of the log-likelihood estimator is greater than one; Ref.~\cite{jack_math_tome} further showed that $\lvar{} < 1$ across the posterior support is a sufficient (though not necessary) condition for unbiased population inference.
Therefore, we adopt $\lvar \approx 1$ as a benchmark when evaluating the tradeoff between uncertainty in the likelihood estimator and population measurement uncertainty.

In Fig.~\ref{fig:joint-var-marginals}, we directly compare the increase in statistical uncertainty and coincident decrease in log-likelihood estimator variance with decreasing catalog size as we increase the detection threshold $\rho_*$.
For a fixed catalog, set of single-event posterior samples, and detectable simulated sources, the variance \lvar{} is a function of the population parameters.
In the rightmost panel of Fig.~\ref{fig:joint-var-marginals}, we show the distribution of \lvar{} over each posterior on $\Lambda$ at each \ac{SNR} threshold.
When we raise the \ac{SNR} threshold to at least $\rho_* = 20$, as satisfied by 221 events, we find that $\lvar \leq 1$ for all posterior samples.
In all other panels of Fig.~\ref{fig:joint-var-marginals}, we show 90\% confidence regions for joint posteriors on the log-likelihood estimator variance and fraction of sources in select domains of primary mass, spin magnitude, and spin tilt.
Here, we reproduce the evolution in statistical uncertainties noted in Sec.~\ref{sec:stat-unc}; our constraints on the structure of the primary mass, spin magnitude, and spin tilt distributions only weaken by $\mathcal{O}(1-10\%)$ up to $\rho_* \sim 15-20$.
Crucially, we could ensure unbiased population inference by raising the \ac{SNR} threshold to $\rho_* = 20$, while our statistical uncertainty on the population distribution of masses and spins at most doubles.

In Fig.~\ref{fig:joint-var-marginals}, we also note that \lvar{} negatively correlates with the fraction of sources with $a_{1,2} > 0.5$, and positively with the fraction of sources with $\cos \theta_{1,2} > 0.98$.
This follows from our choice of a uniform draw distribution in spin magnitudes and cosine spin tilts when estimating $\pdet{}$, which are relatively poor proposal distributions for population models with narrow regions of high density like the beta distribution at small spin magnitudes or the Gaussian component at orbit-aligned spin tilts.
We still recover the true fractions within the 90\% credible regions, indicating that our results at all \ac{SNR} thresholds are not biased by the Monte Carlo integration in the estimate of the population likelihood.
However, we simulated an optimistic scenario where the form of the population model employed during inference was the same as the form of the true astrophysical population.
If we employed a more flexible, higher-dimensional population model, $\lvar{}$ would be larger and our results might become noticeably biased.
This is because weaker models can potentially fit distributions that are highly dissimilar to our choice of draw distribution, reducing the effective sample size.

Finally, in App.~\ref{app:ehat}, we compute the posterior error statistic recently derived by Ref.~\cite{jack_math_tome} and confirm that our analyses do not suffer systematic bias due to Monte Carlo estimation of the population likelihood; further, we confirm that the information loss due to misestimation of the likelihood decreases as the \ac{SNR} threshold is raised.

\subsection{Accuracy requirements} \label{sec:ninj-min}

\begin{figure}
    \centering
    \includegraphics[width=0.99\linewidth]{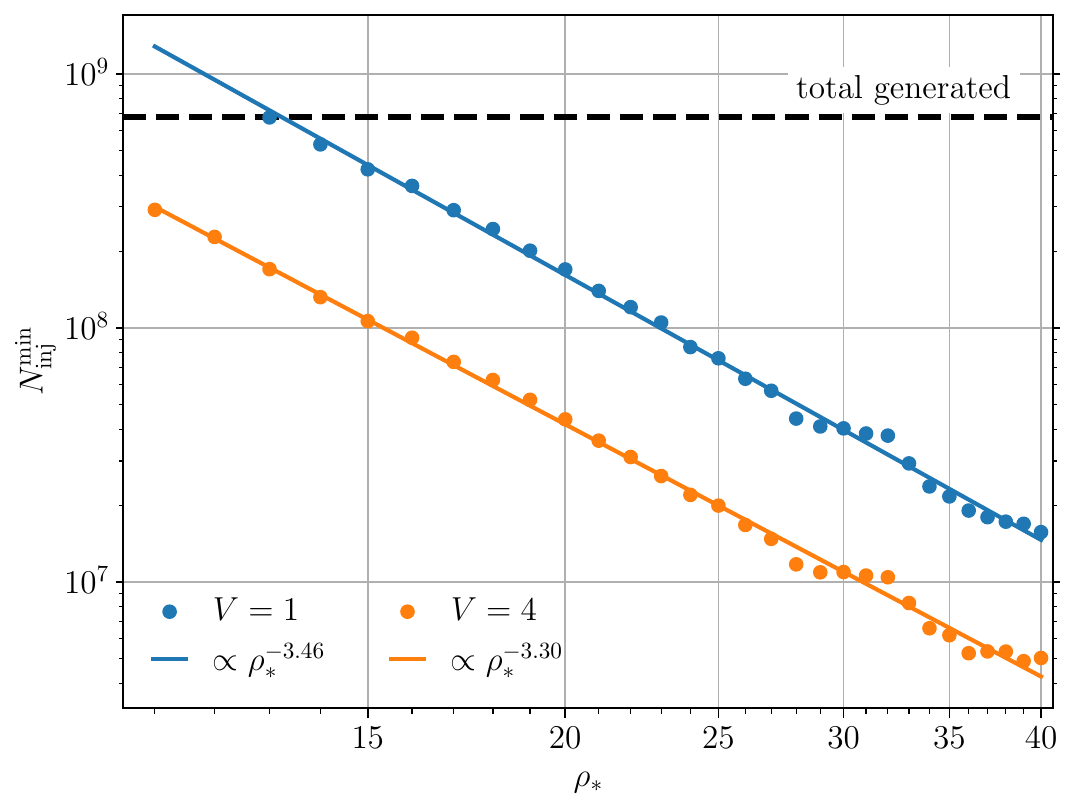}
    \caption{
    The minimum number of simulated sources required for the population log-likelihood estimator variance stemming from selection effects to be less than $V = 1$ (blue dots) or $V = 4$ (orange dots) when evaluated at $\ltrue{}$.
    Power law fits are shown with solid, colored lines.
    In black, we show the total number of simulated sources used for all mock catalog analyses.
    We extrapolate $\nmin{}$ above the total number of sources we simulated, which occurs below $\rho_* = 13$ for $V = 1$.
    }
    \label{fig:min-ninj}
\end{figure}

For the mock catalogs studied in this work, the overall variance \lvar{} of the log-likelihood estimator is dominated by the selection effects term $\vdet{} = \nobs^2 \dvar{}$ (cf. Fig~\ref{fig:vpe-over-vdet-vs-rho}).
However, as we explore in App.~\ref{app:npe}, the relative contribution of selection effects versus the per-event likelihoods to the overall variance scales non-trivially with \ac{SNR} threshold, and $\vdet{}$ dominates, in part, because we drew a large number of samples from each single-event posterior when generating our mock catalog.

We also emphasize that we are only considering Monte Carlo variance of the log-likelihood estimator, not covariance over $\Lambda$ (as in Ref.~\cite{Talbot:2023pex}), nor covariance of the population posterior estimator (as in Refs.~\cite{Essick:2022ojx, jack_math_tome}), which may scale differently than \lvar{} with $\nobs{}$.
While these statistics may provide a better indicator for possible bias due to Monte Carlo integration in population analyses, they have not yet been widely adopted.
Therefore, here we focus on the minimal total number of simulated sources (including detectable and non-detectable) such that the contribution \vdet{} to the overall Monte Carlo variance \lvar{} from estimating the detection efficiency is smaller than a threshold $V$.

Note that \dvar{} may vary with the choice of draw distribution $p(\theta | \ldraw{})$ and the particular set of sources drawn from that distribution---not just their number.
Here, we reuse the set of simulated sources described in Sec.~\ref{sec:methods-mock} and so our choice of draw distribution remains fixed.
We maximize the selection variance over $10^4$ bootstrapped samples (with replacement) of size \ninj{} from \catinj{} to obtain the dependence of the selection variance (at fixed threshold $\rho_*$) $\dvar(\Lambda, N_{\rm inj})$ with \ninj{}.
We define the minimum number of simulated sources required as
\begin{align} \label{eq:nmin-def}
    \nmin{} \left( \Lambda, \nobs{}, V \right) = \min \left\{ \ninj{} : \vdet{} \leq V \right\} \, .
\end{align}
In words, \nmin{} is the fewest number of sources we need to simulate and process with the selection function to keep the variance contribution below $V$, accounting for variability in \catinj{}.
We compute $\nmin{}$ by numerically minimizing $\left( \vdet{} - V \right)^2$ at fixed $\Lambda$.
This optimization is computationally expensive, so we only compute $\nmin{}$ at the true population parameters $\ltrue{}$.

In Fig.~\ref{fig:min-ninj}, we show $\nmin{}$ under two different variance thresholds, $V = 1$ (blue; our benchmark) and $V = 4$ (orange; as adopted in this work).
We fit $\nmin{}$ as a function of $\rho_*$ with a power law, finding that $\nmin{} \propto \rho_*^{-3.46}$ under $V = 1$ and $\propto \rho_*^{-3.30}$ under $V = 4$ (these powers are consistent within uncertainty due to the finite number of $\rho_*$).
In black, we compare $\nmin{}$ to the total number of simulated sources we used to estimate selection effects in our analysis ($\approx 6.4\times10^8$).
At each $\rho_*$ we find that we simulated enough sources to fall below the variance threshold of $V = 4$, in agreement with Fig.~\ref{fig:joint-var-marginals} (recall that Fig.~\ref{fig:joint-var-marginals} shows the variance distributed over each posterior, but in Fig.~\ref{fig:min-ninj} we only compute \dvar{}---and in turn $\nmin{}$---at the true population parameters).
However, more simulations---possibly $\mathcal{O}(10^9)$---would be required to keep for $\dvar{} <1$ when analyzing our largest catalogs selected under $\rho_* = 11, 12$.
While it may be possible to generate and process the required simulations (as we explore in Sec.~\ref{sec:compute}), we can  immediately ensure the accuracy of the population likelihood estimator without additional effort by raising the \ac{SNR} threshold.
We derive an estimate for this scaling relation against which these empirical results can be compared in App.~\ref{app:nmin-scaling}.

\subsection{Computation time} \label{sec:compute}

In Secs.~\ref{sec:tradeoff} and \ref{sec:ninj-min}, we found that at least $\mathcal{O}(10^8)$ simulated sources are required to estimate the detection efficiency $\pdet{}$ with sufficient accuracy, depending on \ac{SNR} threshold and tolerance for Monte Carlo variance.
To evaluate whether sensitivity estimates will remain tractable for catalogs of thousands of events, we provide an order-of-magnitude estimate for the realistic computational cost of analyzing the required simulated signals.

Estimating \pdet{} is typically accomplished by drawing many sources $\theta \sim p(\theta | \ldraw{})$, simulating the \ac{GW} signal from those sources, and adding those signals into detector data where no real signals have been observed \cite{ligo_scientific_collaboration_and_virgo_2023_7890437}.
These data are then filtered by search pipelines, which assign a detection statistic to each simulated source.
To approximate the computational cost of estimating \pdet{}, we first compute the total duration of data to be processed.
Following the procedure defined for the \ac{LVK}'s most recent sensitivity estimation campaign \cite{o4a-injection-campaign}, we assume that signals are generated with a minimum duration of $T_{\rm signal} = 16$\,s and those signals are added into detector data spaced by $\Delta T = 24$\,s.
We need only filter a fraction \fhope{} of sources that pass a ``hopeless cut'' in the optimal \ac{SNR} (cf. Sec.~\ref{sec:methods-mock} in our study and Refs.~\cite{Essick:2023toz, ligo_scientific_collaboration_and_virgo_2023_7890437, o4a-injection-campaign} for analysis of real data): if the optimal \ac{SNR} is low, there is very little probability that noise will scatter the observed power above the threshold for detection.
For the choice of $\ldraw{}$ and optimal \ac{SNR} threshold used in our study, $\fhope{} \sim 40\%$.
This yields a total length of data to be processed of roughly,
\begin{align} \label{eq:ttot}
    T_{\rm tot} = (T_{\rm signal} + \Delta T) \fhope{} N_{\rm inj} \sim 16 N_{\rm inj} \, \mathrm{s}.
\end{align}
Data is processed by search pipelines in fixed-duration files and filtered with a waveform template bank to recover the simulated sources.
When filtering is restricted to times and templates near each simulated source---using only $\mathcal{O}(10^3)$ templates---the search pipeline GstLAL requires $\mathcal{O}(1)$\,min to process a 4096\,s duration file \cite{gstlal-cost}.
Taking GstLAL as a reference, the CPU-time required to process $\fhope{} N_{\rm inj}$ simulated sources is approximately
\begin{align}
    T_{\rm CPU} &\sim \left( \frac{T_{\rm tot}}{ 4096~\rm s} \right) \times1\,\mathrm{min} \label{eq:tcpu} \sim 4\times 10^{-3}~N_{\rm inj}~\,\mathrm{min}
    \, .
\end{align}
For our least stringent \ac{SNR} threshold, we found that we required $N_{\rm inj}\sim10^9$ simulated sources to achieve $\vdet{} \sim 1$.
This translates to $\sim7.6$\,yrs of CPU-time.
Fortunately, this procedure can be parallelized over files and templates; with $\mathcal{O}(10^3)$ CPUs, the wall time is $\mathcal{O}(\mathrm{days})$.\footnote{Our calculation does not include the cost of generating the waveform, only analyzing the data. Waveform generation takes up to $\sim 1$\,s depending on the waveform model (see Ref.~\cite{Pratten:2020ceb}, Tab.~1). This is comparable to the analysis time per-signal if each 4096\,s file has $\sim 100$ signals and requires $\sim 1$\,min to analyze.
}
Note, however, that filtering each file against a full bank of $\mathcal{O}(10^6)$ templates would require a factor of $\mathcal{O}(10^3)$ more CPU-time and wall time, which may be computationally prohibitive when analyzing $\sim 10^9$ simulated sources.

Further, we designed $\ldraw{}$ for a simulated population of BBHs only, with component masses $\geq 3\,M_\odot$.
\ac{GW} signals from lower-mass sources tend to be quieter, reducing \fhope{} and requiring us to simulate additional sources (as \dvar{} scales inversely with the number of detectable sources, see App.~\ref{app:nmin-scaling}).
They are also longer in duration, up to hundreds of seconds in current-generation \ac{GW} detectors.
Thus, accounting for binaries with component masses down to $1\,M_\odot$ or lower as in Refs.~\cite{ligo_scientific_collaboration_and_virgo_2023_7890437, o4a-injection-campaign, LVK:2022ydq} would increase $T_{\rm CPU}$.

\section{Conclusions} \label{sec:discussion}

Population inference can be biased by misspecification or misestimation of the population likelihood.
We estimate the likelihood with Monte Carlo integrals, each of which carry associated variance; when the sum of these variances is large, the population likelihood may incorrectly be large in regions of the population parameter space otherwise unsupported by observed \ac{GW} events.
Accurately estimating the likelihood is computationally expensive, and the accuracy requirements will grow quickly with future \ac{GW} catalogs.
In this work, we evaluated a simple solution to ameliorate these issues: raising the detection threshold.
In particular, we studied the tradeoffs of analyzing
catalogs containing fewer events but with higher \acp{SNR}
in terms of statistical uncertainty, systematic uncertainty, and computational cost.
We summarize our key results in Tab.~\ref{tab:summary}.

\begin{table}[htbp]
\renewcommand*{\arraystretch}{1.1}
\setlength{\tabcolsep}{4pt}
    \centering
    \begin{tabular}{ c | cccc | c | cc }
        \hline
        \hline
        $\rho_*$ & \multicolumn{4}{c|}{ \shortstack{ \relunc{} \\ Sec.~\ref{sec:stat-unc} } } & \shortstack{ $\mathcal{V}$ \\ Sec.~\ref{sec:tradeoff} } & \multicolumn{2}{c}{\shortstack{ \nmin{} [$\times 10^8$] \\ Sec.~\ref{sec:ninj-min}} }  \\
        \cline{2-5}\cline{7-8}
         & $z$ & $m_1$ & $a_{1,2}$ & $\cos\theta_{1,2}$ 
         & & $\mathcal{V} < 1$ & $\mathcal{V} < 4$ \\
        \hline
        11 & -- & --  & -- & -- & 2--3 & $> 6.4$ & 2  \\
        13 & 1.5 & 1.2 & $\gtrsim 1$ & $\gtrsim 1$ & 1--2 & 6.4 & 1.5 \\
        15 & 2 & 1.6 & $\gtrsim 1$ & $\gtrsim 1$ & $\sim 1$ & 4 & 1 \\
        \hline
        \hline
    \end{tabular}
    \caption{
    Summary of analyses performed under \ac{SNR} thresholds $\rho_*$ of 11, 13, and 15.
    From left to right, we show characteristic values for: the growth in 90\% \ac{CI} width \relunc{} on the population distribution relative to our analysis at $\rho_* = 11$ for redshift $z$, primary mass $m_1$, and spin magnitudes $a_{1,2}$ and tilts $\cos \theta_{1,2}$; the variance of the likelihood estimator \lvar{}; and the minimum number \nmin{} of simulations required to bound the selection variance below thresholds of 1 or 4.
    }
    \label{tab:summary}
\end{table}

To study how measurement uncertainty and population likelihood accuracy evolve with the choice of detection threshold, we performed parameter estimation for large catalogs of up to 1600 simulated \ac{GW} signals from \ac{BBH} mergers observed in a network of the LIGO Hanford, LIGO Livingston, and Virgo detectors with current sensitivities.
Detection was determined using the network matched-filter \ac{SNR} evaluated between the noisy simulated data and true \ac{GW} strain.
Source parameters were drawn from a phenomenological distribution without correlations, and the astrophysical distribution of source parameters was fit with that same model in a hierarchical Bayesian manner.

\textbf{The growth in statistical uncertainty with reduced catalog size depends on the source parameters of interest}.
As one might have expected, constraints on the redshift-evolving \ac{BBH} merger rate density were the most sensitive to \ac{SNR} threshold---e.g., increasing the \ac{SNR} threshold from 11 to 13 (in turn reducing \nobs{} from 1600 to 904) increased statistical uncertainty in the merger rate by at most $\sim 50\%$---owing to the inverse relationship between source distance and signal amplitude.
We found that the inferred mass distribution was more robust to increases in \ac{SNR} threshold.
Specifically, the statistical uncertainty in the primary mass distribution increased by $\sim 60\%$ when raising the \ac{SNR} threshold from 11 to 15 (correspondingly reducing \nobs{} to 560).
Uncertainties in the component spin magnitudes and tilt angles remained nearly unchanged even with a tenfold reduction in catalog size.

\textbf{Raising the \ac{SNR} threshold reduces systematic uncertainties.}
We found that increasing the \ac{SNR} threshold from 11 to 15 reduced the Monte Carlo variance of the log-likelihood estimator from $\mathcal{V} \sim 3$ to $\sim 1$.
Further increasing the \ac{SNR} threshold to 20 (with $\nobs = 221$) reduced the likelihood variance below one---ensuring unbiased population inference---while still yielding qualitatively similar conclusions about the astrophysical distributions of \ac{BH} mass and spin.
For large catalogs of \acp{BBH} that we expect to accumulate in the future, the accuracy in the estimate of the population likelihood may be dominated by the selection effects term.
We found that the minimum number of simulated sources needed to ensure accurate estimates of the population log-likelihood scales with the SNR threshold $\rho_*$ steeper than $\rho_*^{-3}$ for our choice of astrophysical population and model.
Estimating selection effects with sufficient accuracy requires processing $\mathcal{O}(10^9)$ simulated signals for catalogs of $\mathcal{O}(10^3)$ events; when selecting based on the response of template-based searches, this procedure should remain tractable with reasonable computational resources when filtering is restricted to templates similar to the simulated signals.

There are additional computational burdens we did not focus on which are influenced by catalog size; namely, the requirement to draw a large number of samples from each single-event posterior.
Drawing single-event posterior samples with nested sampling scales linearly in the number of live points, and in turn, the computational cost per event scales linearly in the desired number of samples \cite{Ashton:2022grj} (we note, however, that the wall time per event can be reduced by combining the outputs of multiple nested sampling routines in parallel).
For \textit{each} source in our mock catalogs, we spent $\gtrsim 1$\,CPU-day
to produce $>15000$ posterior samples for accurate Monte Carlo integration in the population likelihood; significantly more compute time---or resources if parallelizing---could be required, per event, when using more computationally expensive waveform families (see Ref.~\cite{Pratten:2020ceb}).
By raising the \ac{SNR} threshold and increasing the accuracy of the likelihood estimator, we reduce the sampling requirements on single-event parameter estimation.

Our results depend on the astrophysical population and model used to perform population inference.
Our statistical uncertainties---particularly for the spin magnitude and cosine-tilt distributions---are model dependent and would likely broaden if inferred using more flexible population models (see, e.g., Refs.~\cite{
Tiwari:2021yvr, Vitale:2022dpa, Golomb:2022bon, Edelman:2022ydv, Godfrey:2023oxb, Adamcewicz:2023mov, Callister:2023tgi} for studies exploring more flexible spin models).
The estimate of the population log-likelihood estimator would similarly be more uncertain.
Additionally, source properties were not correlated in the astrophysical population we simulated; however, current \ac{GW} datasets suggest that \ac{BH} masses and spins may be correlated \cite{Callister:2021fpo, Biscoveanu:2022qac, Heinzel:2023vkq, Heinzel:2024hva, Pierra:2024fbl, Hussain:2024qzl, Franciolini:2022iaa, Li:2023yyt, Antonini:2025zzw, Sadiq:2025vly, Mould:2022ccw}.
We also knew the true astrophysical distribution and could thus tailor our simulations when estimating selection effects to achieve a high effective sample size and reduce the overall number of simulated sources we needed to process to reduce the uncertainty of the likelihood estimator.
With real data, we do not have this benefit, limiting the degree to which tailoring alone can significantly increase the accuracy of our likelihood estimator.
Further, the degree to which raising the \ac{SNR} threshold reduces Monte Carlo variance from estimating selection effects depends on the astrophysical distribution and choice of simulated-source distribution; our choice of the simulated-source distribution was the most optimistic, and so it is worth studying the reduction in systematic uncertainties at higher detection thresholds while using sensitivity estimates produced in real observing scenarios.
Finally, our mock catalogs were limited to current-generation sensitivity, whereas next-generation detectors will be an order-of-magnitude more sensitive \cite{Evans:2021gyd, Branchesi:2023mws} and their data may come with complications that we ignored, such as overlapping signals \cite{Regimbau:2009rk}.
Further evaluating the balance between the astrophysical fidelity provided by large catalogs and attendant computational complexity with flexible population models in realistic next-generation observing scenarios is critical future work.

Although our numerical experiments are based on relatively uncomplicated population models and simulated data, our results indicate that \ac{GW} populations can be precisely measured even when analyzing only higher-significance events.
Further, that smaller catalogs of high-significance events are less susceptible to systematic biases---from Monte Carlo variance, as studied here, to noise artifacts like glitches---which can be otherwise computationally costly to remedy.
Thus, increasing the detection threshold is a simple and practical remedy for systematic biases and computational expense in \ac{GW} population inference.

\acknowledgments

We thank Sof\'ia \'Alvarez-L\'opez, Sylvia Biscoveanu, Michele Mancarella, Alex Nitz, Cailin Plunkett, Luca Reali, and Javier Roulet for helpful discussions.
We thank Prathamesh Joshi, Reed Essick, and Tito Dal Canton for insights into sensitivity estimation campaigns with \ac{GW} search pipelines.
We thank Asad Hussain for a timely internal review.
N.E.W. and J.H. are supported by the National Science Foundation Graduate Research Fellowship Program under grant No. 2141064.
M.M. is supported by the Royal Commission for the Exhibition of 1851 research fellowship, and was supported by the LIGO Laboratory through the National Science Foundation award and PHY-2309200.
S.V. is partially supported by the NSF grant No. PHY-2045740.
The authors are grateful for computational resources provided by the LIGO Laboratory and supported by National Science Foundation Grants No. PHY-0757058 and No. PHY-0823459.
This material is based upon work supported by NSF's LIGO Laboratory which is a major facility fully funded by the National Science Foundation.
This work is supported by the National Science Foundation under Cooperative Agreement PHY-2019786 (The NSF AI Institute for Artificial Intelligence and Fundamental Interactions, http://iaifi.org/).

\appendix

\section{Scaling of number of simulations with \ac{SNR} threshold} \label{app:nmin-scaling}

The variance $\dvar{}$ of the Monte Carlo estimate of $\ln \pdet{}$ (cf. Eq.~\eqref{eq:monte-carlo-vt}) does not necessarily decrease with \ac{SNR} threshold $\rho_*$.
Here, we derive a scaling relationship between $\dvar{}$ and $\rho_*$,
starting from the effective sample size \neff{} appearing in the Monte Carlo variance of $\ln \pdet{}$.
The effective number of samples \cite{kish1995} drawn from $p(\theta|\Lambda)$ when estimating the detection efficiency \pdet{} using samples drawn from $p(\theta | \ldraw{})$ is
\begin{align}
    \neff &= \frac{ \left( \sum_{i = 1}^{\ninj} w_i \right)^2 }{ \sum_{i = 1}^{\ninj} w_i^2 } = \frac{1}{\sum_{i = 1}^{\ninj} \bar{w}_i^2} \, , \label{eq:neff}
\end{align}
where the weights $w_i$ are
\begin{align}
    w_i(\Lambda) &= \frac{p(\theta_i | \Lambda)}{p(\theta_i | \Lambda_{\rm draw})} \Theta(\rho(d_i) > \rho_*)
\end{align}
and $\bar{w}_i$ are the normalized weights
\begin{align}
    \bar{w}_i = \frac{w_i}{\sum_{i = 1}^{\ninj} w_i} \, .
    \label{eq:norm-weights}
\end{align}
Note, however, that a portion of our weights are exactly zero---in particular, those with $\rho(d_i) \leq \rho_*$.
These weights do not contribute to the normalization $\sum_{i = 1}^{\ninj} w_i$, nor to $\sum_{i = 1}^{\ninj} \bar{w}_i^2$.
Using $k$ to index only the $N_\mathrm{det}$ weights with $\rho(d_k) > \rho_*$, we can write
\begin{align}
    \neff = \frac{1}{\sum_{k = 1}^{\ndet} \bar{w}_k^2} \, , \quad
    \bar{w}_k = \frac{w_k}{\sum_{k = 1}^{\ndet} w_k} \, .
\end{align}
These expressions are equivalent to Eqs.~\eqref{eq:neff} and \eqref{eq:norm-weights}, respectively.
Importantly, we still have that $\bar{w}_k$ are normalized, with $\sum_{k = 1}^{\ndet} \bar{w}_k = 1$.

\begin{figure}[h]
    \centering
    \includegraphics[width=0.8\linewidth]{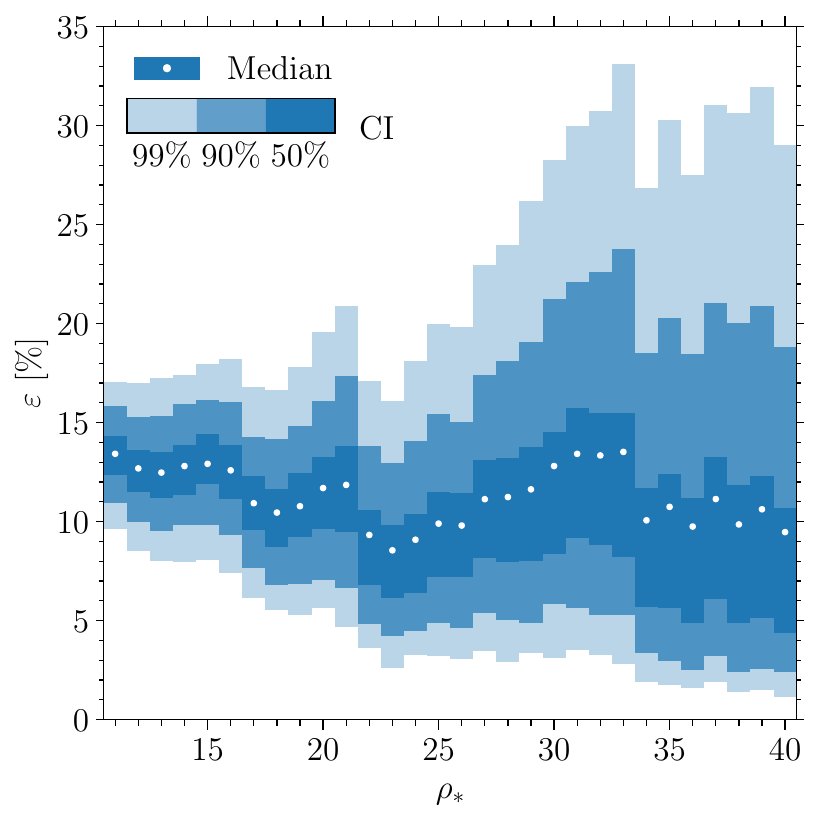}
    \caption{Importance sampling efficiency $\varepsilon$ from the distribution of simulated sources to population distributions parametrized by $\Lambda$ over the posterior support of $\Lambda$ as a function of \ac{SNR} threshold $\rho_*$.}
    \label{fig:eff-vs-rho}
\end{figure}

We recall the Cauchy--Schwarz inequality in a Euclidean $\mathbb{R}^{n}$ vector space (equipped with the standard inner product).
For two vectors $[ u_1, u_2, \ldots, u_n ]^\intercal$ and $[ r_1, r_2, \ldots, r_n ]^\intercal$, we have
\begin{align}
    \left( \sum_{k = 1}^n  u_k r_k \right)^2 \leq \left( \sum_{k = 1}^n u_k^2 \right) \left( \sum_{k = 1}^n r_k^2 \right) \, .
\end{align}
With $n = \ndet$, $u_k = \bar{w}_k$, and $r_k = 1$, this implies
\begin{align}
    \left( \sum_{k = 1}^{\ndet}  \bar{w}_k \right)^2 \leq \ndet \sum_{k = 1}^{\ndet} \bar{w}_k^2 \, .
\end{align}
Recall that $\bar{w}_k$ are normalized, so we have
\begin{align}
    \frac{1}{\ndet} \leq \sum_{k = 1}^{\ndet} \bar{w}_k^2 \, ,
\end{align}
which implies that $\neff \leq \ndet$.
Therefore, we can write $\neff = \varepsilon \ndet$, where $0 \leq \varepsilon \leq 1$ and Eq.~\eqref{eq:vdet} can be written
\begin{align} \label{eq:vdet-vs-ndet}
    \dvar = \frac{1}{\varepsilon \ndet} - \frac{1}{\ninj} \, .
\end{align}
If the set of simulated sources is fixed,
then $\dvar{} \propto \ndet^{-1}$ and $\vdet{} \propto \nobs^2 / \ndet \propto \rho_*^{b - 2a}$,
assuming $\nobs{} \propto \rho_*^{-a}$, $\ndet{} \propto \rho_*^{-b}$ for some powers $a$ and $b$ \cite{Schutz:2011tw, Chen:2014yla}.
This also assumes that $\varepsilon$ does not change with \ac{SNR} threshold;
in Fig.~\ref{fig:eff-vs-rho} we see that the median $\varepsilon$ over the posterior support only decreases by at most a few percent as the \ac{SNR} threshold increased.

In Sec.~\ref{sec:ninj-min} we found an empirical scaling of the number of simulations \nmin{} required for accurate estimation of \pdet{} with \ac{SNR} threshold $\rho_*$.
We now check our results by deriving a scaling relationship between \nmin{} and $\rho_*$,
starting from Eq.~\ref{eq:vdet-vs-ndet}.
By definition, $\ndet = \pdet(\ldraw) \ninj$, so (suppressing the dependence of \pdet{} on \ldraw{}),
\begin{align}
    \dvar = \frac{1}{\ninj} \left( \frac{1}{\varepsilon \pdet} - 1 \right) \, .
\end{align}
If a small fraction of sources from the draw population are detectable (which is true even under our least stringent \ac{SNR} threshold of 11, where $\pdet \approx 1.6 \times 10^{-2}$), then $\dvar \approx 1 / \left( \varepsilon \pdet \ninj \right)$.
With the definition of $\nmin{}$ in Eq.~\eqref{eq:nmin-def}, we have $V \approx N_\mathrm{obs}^2 / \left( \varepsilon \pdet \nmin \right)$.

Assuming that $\varepsilon$ does not significantly evolve with \ac{SNR},
that the number of detected sources that we expect in the true universe will scale with \ac{SNR} threshold like $\propto \rho_*^{-a}$
and similarly that the number of detected sources in the $\ldraw{}$ universe scales like $\propto \rho_*^{-b}$,
we have
\begin{align} \label{eq:nmin-prop-snr}
    \nmin{} \propto \nobs^2 \pdet^{-1} \propto \rho_*^{b - 2a}.
\end{align}
In the most idealized case that $\ldraw{} = \ltrue{}$ and further that $\ltrue{}$ describes a flat universe with a merger rate constant in luminosity distance and uncorrelated with other source properties,
$a = b = 3$ \cite{Schutz:2011tw, Chen:2014yla} and so $\nmin{} \propto \rho_*^{-3}$.
In this work, we estimate that $a \approx 3.21$ for the true astrophysical distribution and $b \approx 3.25$ for the $\ldraw{}$ distribution (cf. Tab.~\ref{tab:nobs}), yielding $\nmin{} \propto \rho_*^{-3.17}$ (which differs from $\rho_*^{-3}$ because we simulated a non-flat universe with a merger rate that increases with distance).
This predicted scaling differs slightly from the scaling relationships we numerically measured in Fig.~\ref{fig:min-ninj}, likely reflecting a nontrivial scaling of the importance sampling efficiency from $p(\theta | \ldraw{})$ to $p(\theta | \ltrue{})$ with \ac{SNR} threshold (cf. Fig.~\ref{fig:eff-vs-rho}).

\section{Population parameter values} \label{app:pop-pars}

In Tab.~\ref{tab:true-pop}, we record the parameters $\Lambda^\mathrm{true}$ specifying the true astrophysical distribution that we drew sources from to construct our mock catalog; see Sec.~\ref{sec:methods-mock} for the form of the distribution.
We also record priors, which are the same for all analyses in this work.

When drawing sources to estimate selection effects we use the same phenomenological form for the redshift, primary-mass, and mass-ratio distributions as the true astrophysical population.
For the mass parameters in $\ldraw{}$, we take $m_{\min} = 2.4\,M_\odot$, $m_{\max} = 150\,M_\odot$, and $\delta_m = 10\,M_\odot$ to ensure that $\mathrm{supp}~p(\theta | \Lambda) \subset \mathrm{supp}~p(\theta | \ldraw{})$ for all $\Lambda$ in the support of the prior.
Otherwise, we adopt the true parameters of the mass distribution when drawing sources to estimate selection effects.
Similarly, we adopt the same $\lambda_z$ as in the true population.
We draw spin magnitudes and cosine spin tilts from uniform distributions on their physical domains of $[0, 1]$ and $[-1, 1]$, respectively.
We draw sky locations from a uniform distribution on the sky and draw the relative orientation of each source from an isotropic distribution.
The coalescence time at geocenter is drawn from a discrete uniform distribution of times within a 24-hour period.

\begin{table}[h]
    \centering
    \begin{tabular}{ c c c c }
        \hline
        \hline
        \textbf{Parameter} & \textbf{Symbol} & \textbf{Value} & \textbf{Prior} \\
        \hline
        $m_1$ power law index & \mpli{} & 3.4 & $[-1, 15]$ \\
        $q$ power law index & $\beta$ & 1.1 & $[-2, 7]$ \\
        Peak mean & $\mu_m$ & $34\,M_\odot$ & $[20, 50]\,M_\odot$ \\
        Peak standard deviation & $\sigma_m$ & $3.6\,M_\odot$ & $[1, 15]\,M_\odot$ \\
        Peak mixing fraction & \mgbr{} & 0.04 & $[0, 1]$ \\
        Minimum \ac{BH} mass & $m_{\min{}}$ & $5\,M_\odot$ & $[3, 10]\,M_\odot$ \\
        Maximum \ac{BH} mass & $m_{\max{}}$ & $87\,M_\odot$ &  $[70, 150]\,M_\odot$ \\
        Low-mass smoothing scale & $\delta_m$ & $4.8\,M_\odot$ & $[0, 20]\,M_\odot$ \\
        \hline
        Mean & $\mu_\chi$ & 0.3 & $[0, 1]$ \\
        Squared standard deviation & $\sigma_\chi^2$ & 0.04 & $[0.005, 0.1]$ \\
        \hline
        Peak standard deviation & $\sigma_{\cos \theta}$ & 1.18 & $[0.4, 5]$ \\
        Peak mixing fraction & $\xi_{\cos \theta}$ & 0.97 & $[0, 1]$ \\
        \hline
        $z$ power law index & $\lambda_z$ & 2.73 & $[-10, 10]$ \\
        \hline
        \hline
    \end{tabular}
    \caption{
    Population parameters $\ltrue{}$ specifying the astrophysical distribution of masses (top), spin magnitudes (second from top), spin tilts (second from bottom), and redshifts (bottom).
    The beta distribution we adopt for spin magnitudes is nonsingular, so the alpha and beta parameters of that distribution are greater than one.
    We also show priors on each parameter used during population inference; these are the same for all mock catalog analyses.
    We denote a uniform prior between $a$ and $b$ as $[a, b]$.    
    }
    \label{tab:true-pop}
\end{table}

\section{Setup for waveform generation and single-event parameter inference} \label{app:single-event-settings}

Throughout our work, we use the same settings to simulate signals from sources in our mock catalog and when simulating signals to estimate selection effects.
We simulate signals using the frequency-domain waveform approximant \textsc{IMRPhenomXP} \cite{Pratten:2020ceb} between a minimum frequency $f_{\min{}} = 20\,\rm Hz$ and maximum frequency $f_{\max{}} = 1024\,\rm Hz$.
We define the component BH spin vectors and binary angular momentum at a reference frequency of $20\,\rm Hz$.
We set \texttt{PhenomXPrecVersion} to \texttt{`104'} (Ref.~\cite{Pratten:2020ceb}, Tab.~III).
To set the duration of the simulated signal, we use a post-Newtonian approximation to the inspiral duration and round that value up to the nearest power of 2 that is at least $16\,\rm s$.
We heterodyne the single-event likelihood using the true \ac{GW} strain as the reference waveform to speed up parameter estimation \cite{Cornish:2010kf, Zackay:2018qdy, Cornish:2021lje, Krishna:2023bug}.
We set the maximum dephasing between frequency bins to be $\epsilon = 0.025$ (see Ref.~\cite{Krishna:2023bug}).

\begin{table}[b]
    \centering
    \begin{tabular}{l l}
        \hline
        \hline
        \textbf{Parameter} & \textbf{Prior} \\
        \hline
        Chirp mass $\mathcal{M}$ &  $[\mathcal{M}^{\rm true} - \frac{1}{2}\Delta\mathcal{M},\,\mathcal{M}^{\rm true} + \frac{1}{2}\Delta\mathcal{M}]$ \\
        Mass ratio & [0.125, 1] \\
        Spin magnitudes & Uniform \\ 
        Spin orientations & Isotropic \\
        \hline
        Luminosity distance $D_L$ & $[\frac{1}{4}D_L^{\rm true}, 4.2D_L^{\rm true}]$ \\
        Right ascension & Uniform \\
        Declination & Cosine \\
        Inclination angle & Isotropic \\
        Polarization angle & Uniform \\
        Coalescence phase & Uniform \\
        Coalescence time $t_c$ & $[t_c^{\rm true} - 0.1\,\mathrm{s},\,t_c^{\rm true} + 0.1\,\mathrm{s}]$ \\
        \hline
        \hline
    \end{tabular}
    \caption{
    Priors used during parameter estimation for sources in our mock catalog, organized between intrinsic (top) and extrinsic (bottom) parameters.
    ``Uniform" distributions are uniform over the relevant physical domain.
    Otherwise, we denote a uniform prior between $a$ and $b$ as $[a, b]$.
    True parameter values of the simulated source are denoted with a superscript ``true''.
    We take $\Delta \mathcal{M} = 10^{-2}(\mathcal{M}^{\rm true})^{5/2}$.
    The chirp mass $\mathcal{M}$ is defined in the detector frame and the coalescence time $t_c$ is geocenteric.
    For reference on typical priors, see App.~B of Ref.~\cite{Romero-Shaw:2020owr}.
    }
    \label{tab:single-event-priors}
\end{table}

We summarize the single-event priors adopted for parameter estimation on each simulated source in our mock catalog in Tab.~\ref{tab:single-event-priors}.
Most of these are standard choices; however, since we know the true values of our mock sources, we can tailor our priors to speed up parameter estimation.
First, we center our prior on the detector-frame chirp mass $\mathcal{M}$ at the true value $\mathcal{M}^{\rm true}$ for each event.
We determine the width $\Delta \mathcal{M}$ of this prior as a function of the true chirp mass.
Heuristically, sources with smaller chirp masses may be observed for longer and have relatively well-measured chirp masses by accumulating observed phase evolution in the signal.
We select $\Delta \mathcal{M}$ as a function of $\mathcal{M}^{\rm true}$ by eye using the events simulated in Ref.~\cite{Vitale:2025lms}, adopting $\Delta \mathcal{M} = 10^{-2} \left( \mathcal{M}^{\rm true} \right)^{5/2}$.
We also enforce that the minimum of the prior on $\mathcal{M}$ does not go below $1\,M_\odot$ and that the maximum does not go above $200\,M_\odot$.
Similarly, we choose prior bounds in luminosity distance $D_L$ based on the true luminosity distance $D_L^{\rm true}$ and enforce that the minimum of the prior does not go below 1\,Mpc.
The scaling of the minimum and maximum $D_L$ prior bounds shown in Tab.~\ref{tab:single-event-priors} is the same as used in Ref.~\cite{Vitale:2025lms}.
We manually inspect the marginal posteriors on $\mathcal{M}$ and $D_L$ obtained for each event, widening the priors and rerunning parameter estimation for those with noticeable prior railing.

\section{Population parameter uncertainties} \label{app:hyperpars}

In this work, we used a simple phenomenological model with 13 parameters to fit the population of BBH implied by our mock observations;
we chose to use the same form of the astrophysical population model that our mock sources were drawn from.
Here, we record the posteriors on each population parameter for our analysis of each mock catalog selected under \ac{SNR} thresholds $\rho_* \in \{ 11, \ldots, 40 \}$.

\subsection{Mass population parameters}

\begin{figure}[h]
    \centering
    \includegraphics[width=0.99\linewidth]{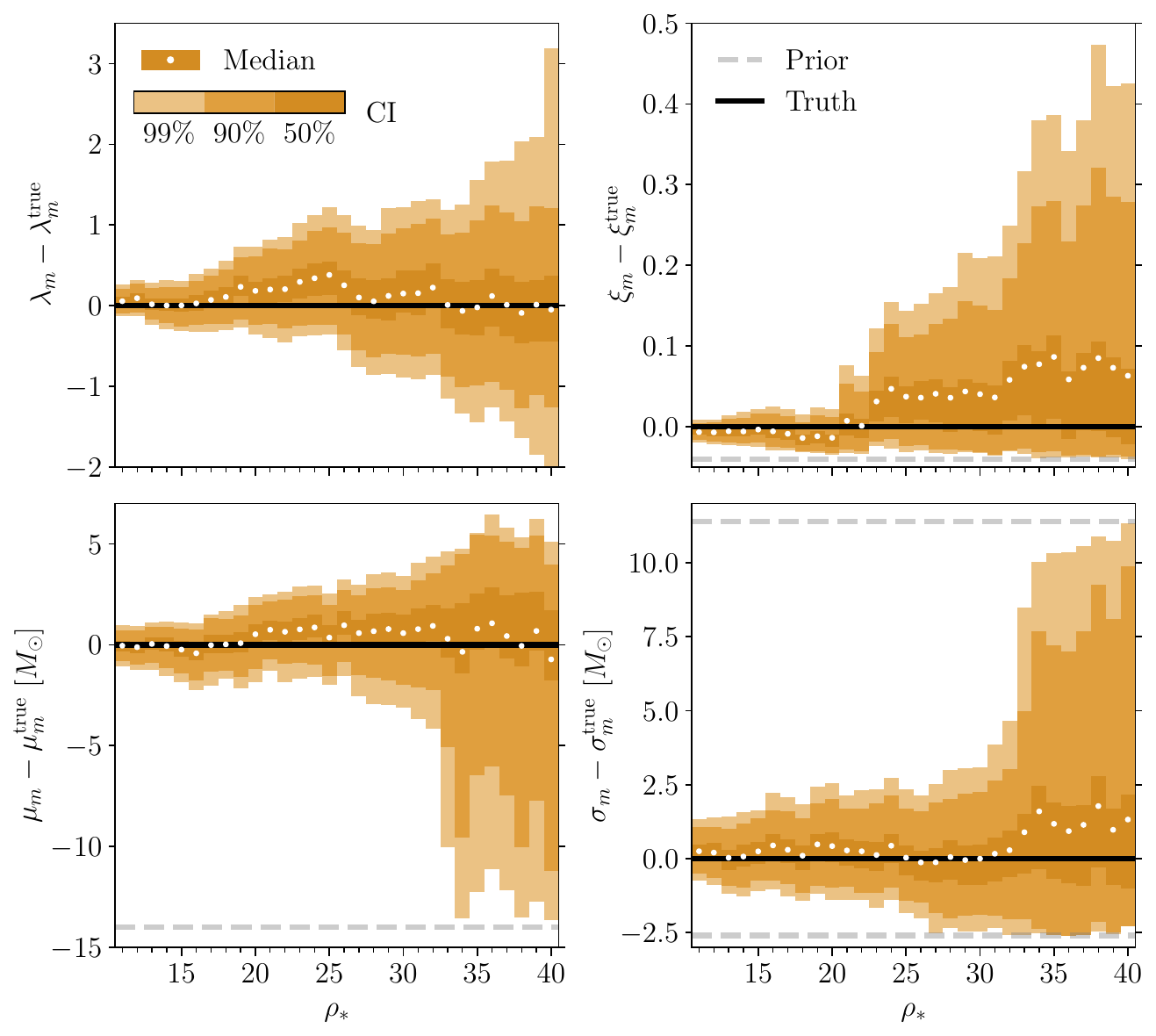}
    \caption{
    Marginal posteriors on four parameters of the primary-mass distribution with respect to their true values: index $-\mpli$ of the power-law continuum (top left), fraction \mgbr{} of sources in the Gaussian peak (top right), which is centered at $\mu_m$ (bottom left) and has standard deviation $\sigma_m$ (bottom right).
    White circles denote medians, while shades from darker to lighter denote the 99\%, 90\%, and 50\% \acp{CI}.
    A black line denotes the truth.
    Prior boundaries are shown with gray dashed lines.
    }
    \label{fig:mock-boxy-alpha-lam-mpp-sigpp}
\end{figure}

\begin{figure}[h]
    \centering
    \includegraphics[width=0.99\linewidth]{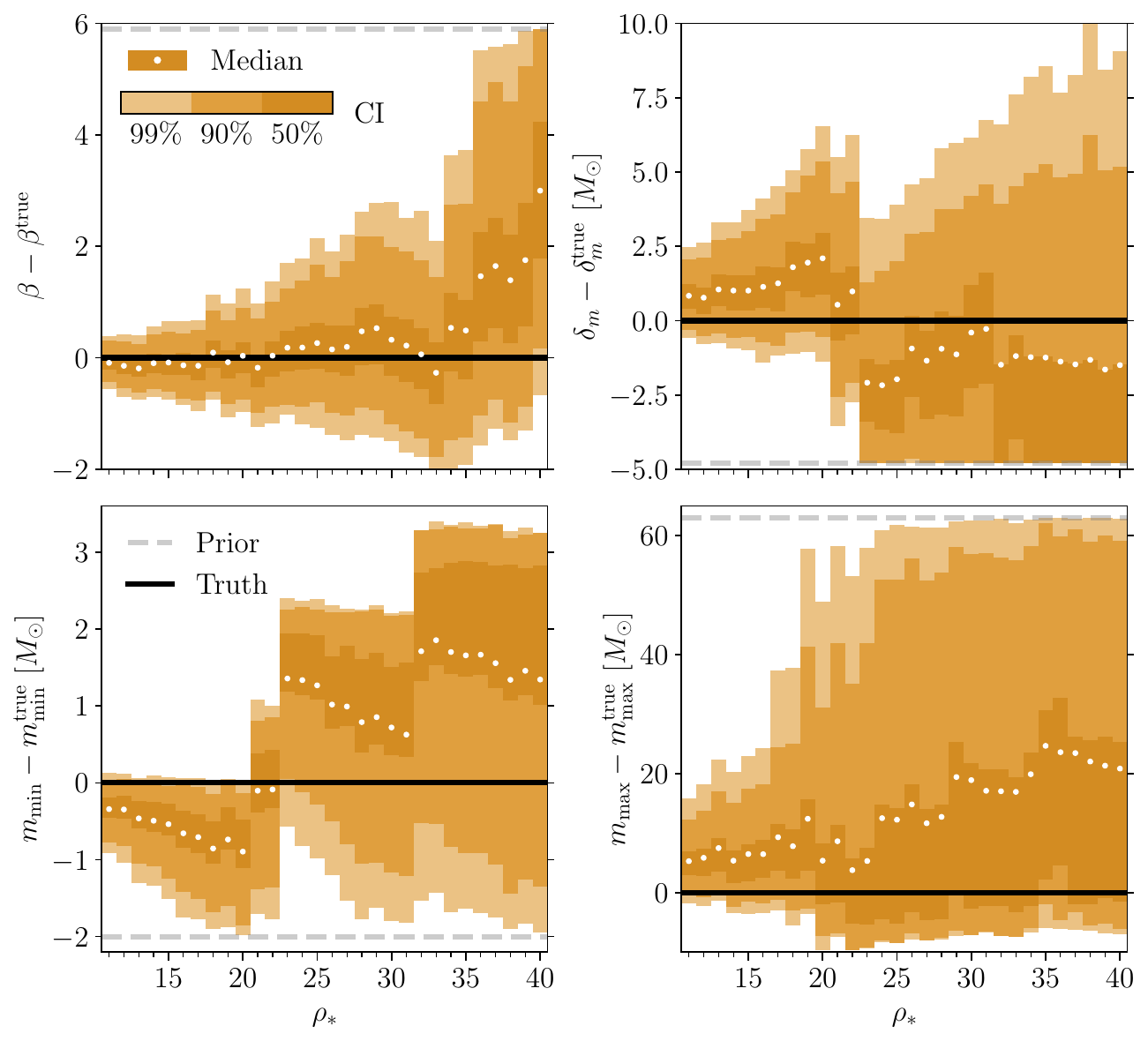}
    \caption{
    Marginal posteriors on four parameters of the primary-mass and mass-ratio distributions with respect to their true parameters: index $\beta$ of the mass-ratio power law (top left), low-mass smoothing scale $\delta_m$ (top right), minimum \ac{BH} mass $m_\mathrm{min}$ (bottom left), and the maximum \ac{BH} mass $m_{\max}$.
    Format matches Fig.~\ref{fig:mock-boxy-alpha-lam-mpp-sigpp}.
    Note that $\delta_m$ rails on the lower prior boundary at zero for $\rho_* \geq 23$.
    }
    \label{fig:mock-boxy-beta-deltam-mmin-mmax}
\end{figure}

In Figs.~\ref{fig:mock-boxy-alpha-lam-mpp-sigpp} and \ref{fig:mock-boxy-beta-deltam-mmin-mmax}, we show marginal posteriors on the population parameters (offset by their true values) for the mass distribution.
In particular, we show the 50\%, 90\%, and 99\% \acp{CI}.
First, we correctly recover the true mass population parameters within at least the 99\% credible regions in all analyses.
Next, we recall from Sec.~\ref{sec:primary-mass} that our constraints on the mass distributions---at all masses as well as integrating over particular domains of primary mass $m_1$---were only moderately relaxed even when the catalog size was reduced by a factor of three or more when increasing $\rho_*$ from 11 to 15 or 20.
The evolution of our constraints on the $m_1$ distribution with catalog size (via changing \ac{SNR} threshold) reflects the moderate increase in statistical uncertainty in the mass population parameters as $\rho_*$ increases.
For example, when $\rho_*$ is increased from 11 to 15---and in turn the catalog size of 1600 is reduced by a factor of nearly three---the 90\% \ac{CI} on $\mpli{}$ grows by a factor of $\sim1.6$.
When $\rho_*$ is further increased to 20, as satisfied by only 221 events, our uncertainty on $\mpli{}$ is a factor of 2.7 larger than at $\rho_* = 11$ (cf. top-left panel of Fig.~\ref{fig:mock-boxy-alpha-lam-mpp-sigpp}).

Similarly, fluctuations in the low-mass behavior of the mass distribution reflect Poisson noise in $m_{\min}$ and $\delta_m$, the minimum BH mass and the smoothing scale of the low-mass tail of the mass distribution, respectively.
These parameters (plus the index of the power-law continuum \mpli{}, although this is constrained by BBH at many masses) most directly control the shape of the population distribution at low masses $\lesssim 7\,M_\odot$.
In the top-right and bottom-left panels of Fig.~\ref{fig:mock-boxy-beta-deltam-mmin-mmax}, the inferred values and constraints on $m_{\min}$ and $\delta_m$ appear to change sharply when $\rho_*$ is increased from $\rho_* = 20$ to $\rho_* = 21$.
This is the result of Poisson noise in catalog membership, as the tails of the mass distribution are most strongly determined by the extremal draws and there are relatively few detectable sources (at all \ac{SNR} thresholds) with component masses $\sim 3\,M_\odot$ in our astrophysical population.

\subsection{Spin population parameters}

\begin{figure}[h]
    \centering
    \includegraphics[width=0.99\linewidth]{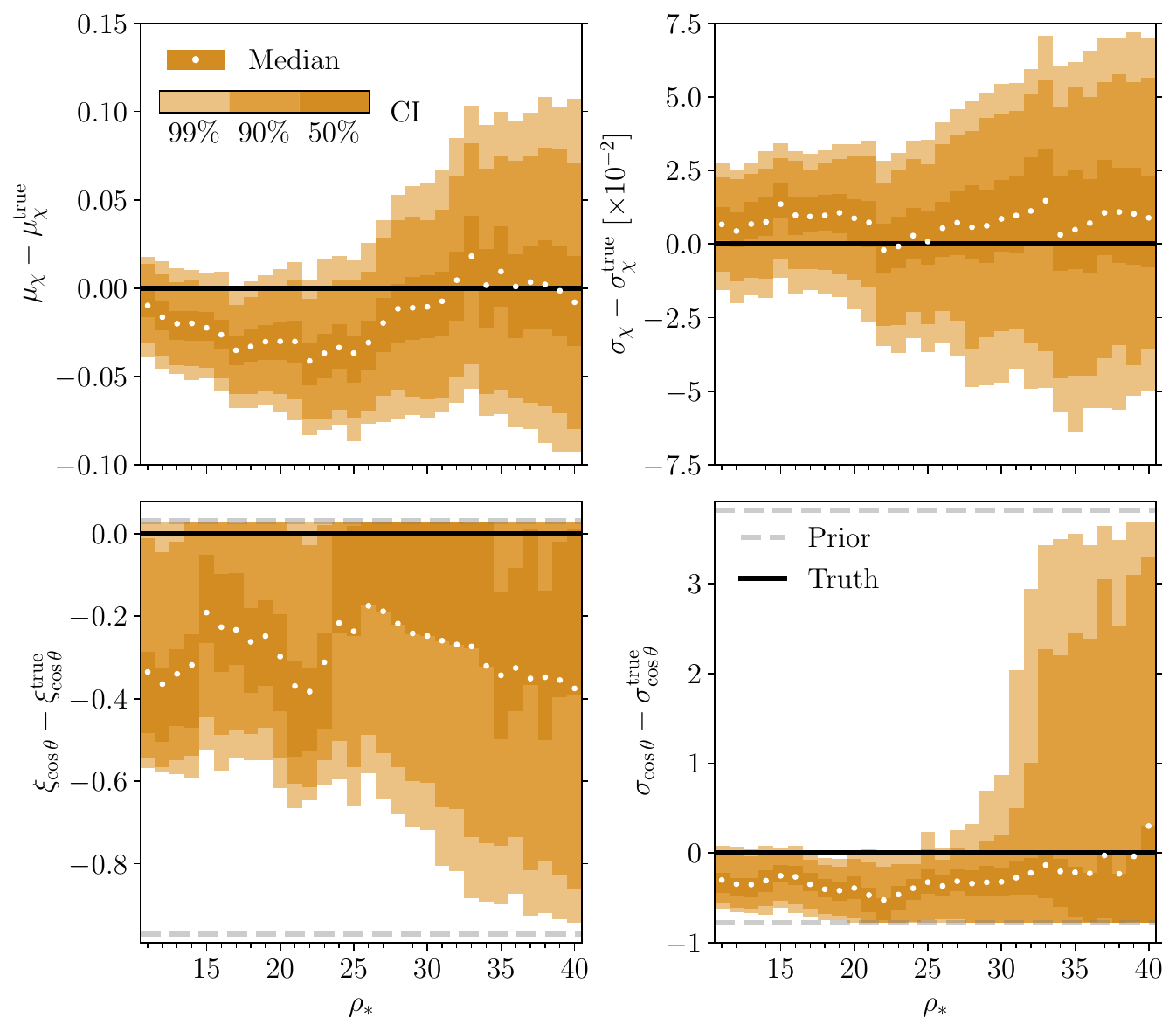}
    \caption{
    Marginal posteriors on the mean $\mu_\chi$ and standard deviation $\sigma_\chi$ of the spin-magnitude distribution, and the standard deviation $\sigma_{\cos \theta}$ and mixing fraction $\xi_{\cos \theta}$ of the Gaussian component of the spin-tilt distribution.
    Format matches Fig.~\ref{fig:mock-boxy-alpha-lam-mpp-sigpp}.
    }
    \label{fig:mock-boxy-spin}
\end{figure}

In Fig.~\ref{fig:mock-boxy-spin}, we show posterior \acp{CI} for marginal posteriors on the parameters of the spin-magnitude and spin-tilt distributions.
The true values of the mean $\mu_\chi$ and standard deviation $\sigma_\chi$ of the spin-magnitude distribution, as well as the true mixing fraction $\xi_{\cos \theta}$ of the Gaussian component of the spin-tilt distribution, are all recovered within, at most, the 90--99\% \acp{CI} in all analyses.
The standard deviation of the Gaussian component of the spin-tilt distribution $\sigma_{\cos \theta}$ is only recovered within the 99.9\% \ac{CI} in our analyses at $\rho_* = 18, 19$, but we found no cause for this apparent bias.
Generally, our constraints on the spin population parameters are essentially unchanged up to $\rho_* \sim 25$, apart from some scatter due to catalog membership.

\subsection{Redshift power-law index}

\begin{figure}[h]
    \centering
    \includegraphics[width=0.85\linewidth]{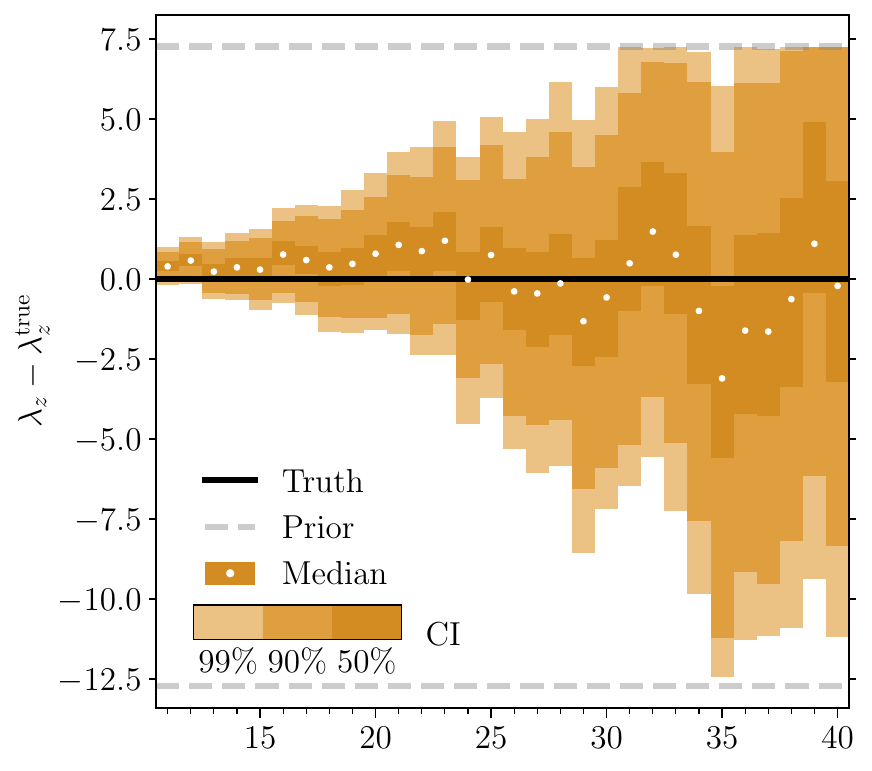}
    \caption{
    Marginal posteriors on the power-law index $\lambda_z$ of the redshift distribution.
    Format matches Fig.~\ref{fig:mock-boxy-alpha-lam-mpp-sigpp}.
    We note that $\lambda_z$ rails on the upper prior boundary at 10 for most analyses with $\rho_* \gtrsim 30$.
    }
    \label{fig:mock-boxy-lamb}
\end{figure}

In Fig.~\ref{fig:mock-boxy-lamb} we show \acp{CI} for the marginal posterior on the power-law index of the redshift distribution, $\lambda_z$.
Here, we note that our statistical uncertainty on $\lambda_z$ grows more quickly with increasing \ac{SNR} threshold $\rho_*$ than our uncertainty in the mass or spin population parameters.
This is because higher redshift sources have lower \acp{SNR}, the distribution of which strongly peaks at the \ac{SNR} threshold.
When $\rho_*$ increases from 11 to 15, our uncertainty on $\lambda_z$ grows by a factor of $\approx2.2$; when $\rho_*$ is further increased to $\rho_* = 20$, our uncertainty is a factor of $\approx4.4$ larger than when analyzing all 1600 events with $\rho > 11$.

\section{Catalog size versus composition} \label{app:nobs69-additional}

In Sec.~\ref{sec:stat-unc}, we observed that constraints on the marginal population-level distribution of \ac{BBH} redshifts were more sensitive to the \ac{SNR} threshold than was the case for primary mass.
We also found that constraints on the spin magnitude and tilt distributions only noticeably weakened when the \ac{SNR} threshold was raised to $\rho_* \sim 20$.
These results indicate that population inference under our choice of population model is driven---at least in part---by the particular events composing our mock catalog, and not only the size of the catalog.
Here, we directly assess whether catalog size or composition informs the population distribution by repeating population inference on many catalogs of fixed size while varying \ac{SNR} threshold.

\begin{figure}
    \centering
    \includegraphics[width=0.95\linewidth]{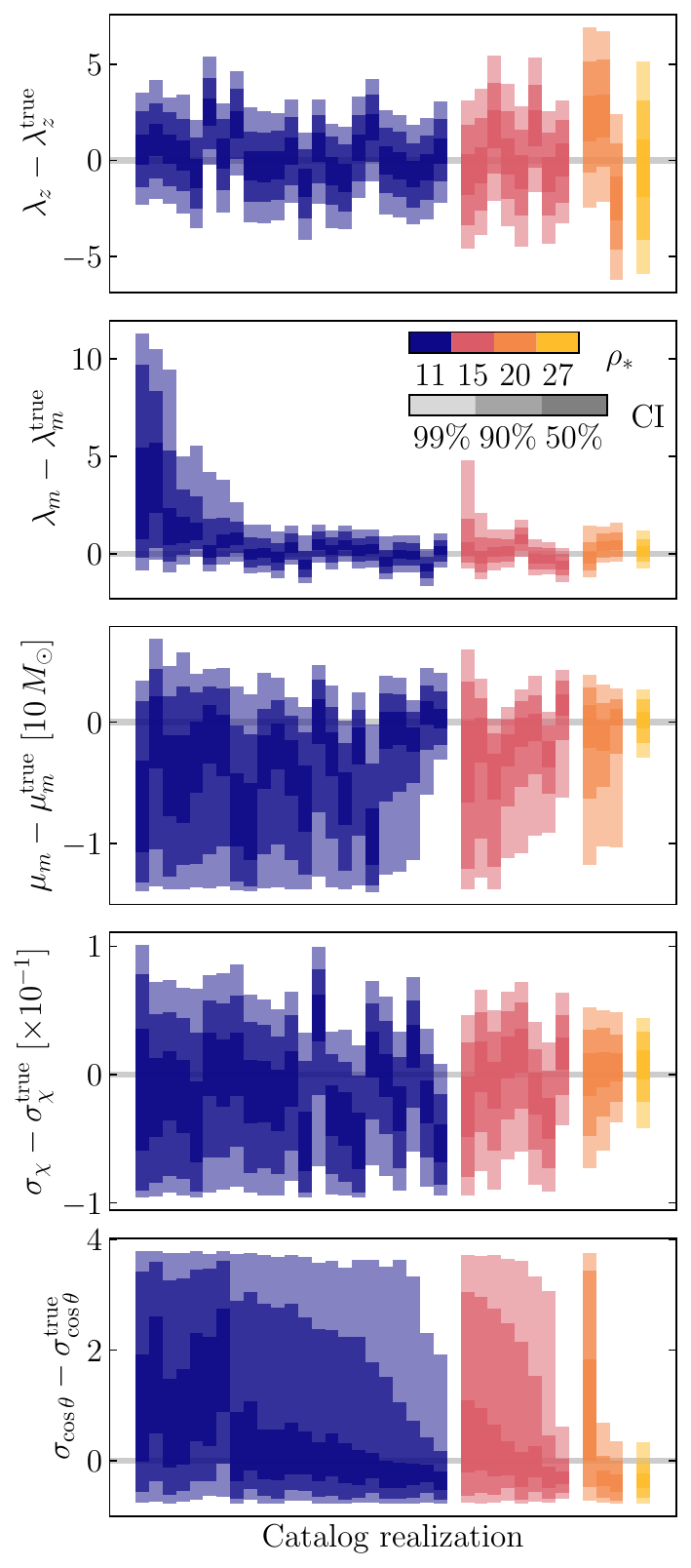}
    \caption{
    Marginal posteriors from analyses of unique subsets of 69 events from our mock catalog. Rows from top to bottom show the results for the power-law index $\lambda_z$ of the redshift distribution, power-law index $-\mpli{}$ of the continuum in primary mass, location $\mu_m$ of the Gaussian peak in primary mass, standard deviation $\sigma_\chi$ of the spin-magnitude distribution, and standard deviation $\sigma_{\cos \theta}$ of the Gaussian component of the spin-tilt distribution.
    Colors indicate different \ac{SNR} thresholds and shades from darker to lighter denote the 99\%, 90\%, and 50\% \acp{CI}.
    At each $\rho_*$, posteriors are sorted by decreasing width of the 90\% \ac{CI}.
    }
    \label{fig:nobs69-mock-boxy}
\end{figure}

First, we perform hierarchical inference on mock catalogs of fixed size and increasing \ac{SNR} threshold.
We take unique subsets of $\nobs = 69$ from our mock catalog of 1600 events at four thresholds $\rho_* \in \{ 11, 15, 20, 27 \}$
(chosen from among the thresholds highlighted in the previous section, replacing $\rho_* = 30$ with $\rho_* = 27$, the highest threshold satisfied by at least 69 events).
We note that we have only one catalog of 69 events with $\rho > 27$, and so cannot test how our results vary with different catalog realizations selected with $\rho_* = 27$.

We summarize our results in Fig.~\ref{fig:nobs69-mock-boxy}, where we show posterior \acp{CI} of some of the parameters characterizing the distributions of redshift, primary mass, spin magnitude, and spin tilt, relative to their true values.
Beginning from the top, we find that catalogs of 69 events with $\rho > 11$ typically yield the best measurement of the power-law index $\lambda_z$ of the redshift distribution, with a trend of increasing statistical uncertainty as $\rho_*$ increases.
The smallest 90\% \ac{CI} width for $\lambda_z$ achieved among the catalogs selected under $\rho_* = 11$, 15, 20 and 27 are 3.2, 4.3, 5.5, and 7.3, respectively.
Therefore, the redshift distribution is better informed by including events with lower \ac{SNR}.
Events with lower \ac{SNR} tend to lie at further distances, thus providing a crucial lever arm to constrain the evolution of the merger rate over redshift.

In the next row, we turn to the index \mpli{} of the power-law continuum in primary mass $m_1$.
Here, the smallest 90\% \ac{CI} widths among catalogs at each $\rho_*$ are 1.1, 1.1, 1.3, and 1.2.
These results indicate that the broad shape of the mass distribution is not affected by the \ac{SNR} threshold.
Conversely, the location $\mu_m$ of the Gaussian peak in $m_1$ is best measured with a catalog of relatively high \ac{SNR} events; among catalogs selected with $\rho_* = 11$, 15, 20, and 27, the smallest 90\% \ac{CI} widths are $4.4\,M_\odot$, $4.8\,M_\odot$, $4.5\,M_\odot$, and $3.6\,M_\odot$, respectively.

Finally, we turn to the standard deviation of the spin magnitude distribution, $\sigma_\chi$, and the standard deviation of the Gaussian component in spin tilts, $\sigma_{\cos \theta}$.
Both marginals show a trend of decreasing uncertainty as $\rho_*$ is raised.
Among catalogs selected with \ac{SNR} thresholds $\rho_*=11$, 15, 20 and 27, the smallest 90\% \acp{CI} respectively are 0.070, 0.063, 0.052, and 0.055 for $\sigma_\chi$ and 0.9, 0.6, 0.73, and 0.63 for $\sigma_{\cos \theta}$.
These results indicate that spin population inference is driven by events with higher \ac{SNR}.
Additionally striking is the variation in statistical uncertainty under different catalog realizations when measuring $\sigma_{\cos \theta}$; up to $\rho_* = 20$, many catalogs yield statistical uncertainty on $\sigma_{\cos \theta}$ comparable to the width of the prior.
This variability suggests that additional characteristics of the sources in our mock catalogs---beyond \ac{SNR}---determine how well the spin tilt distribution can be constrained.

While population inference is ultimately determined by the data in a \ac{GW} catalog, these results indicate that we can crudely dissect the information content of our catalogs based on size and event \acp{SNR}.
In a future study, we will further dissect the overall information content of our mock catalogs in terms of the information content in each event and the (inferred) parameters of each \ac{BBH}.

\section{Scaling of Monte Carlo variance with \npe{} and \ac{SNR} threshold} \label{app:npe}

\begin{figure}[h]
    \centering
    \includegraphics[width=0.85\linewidth]{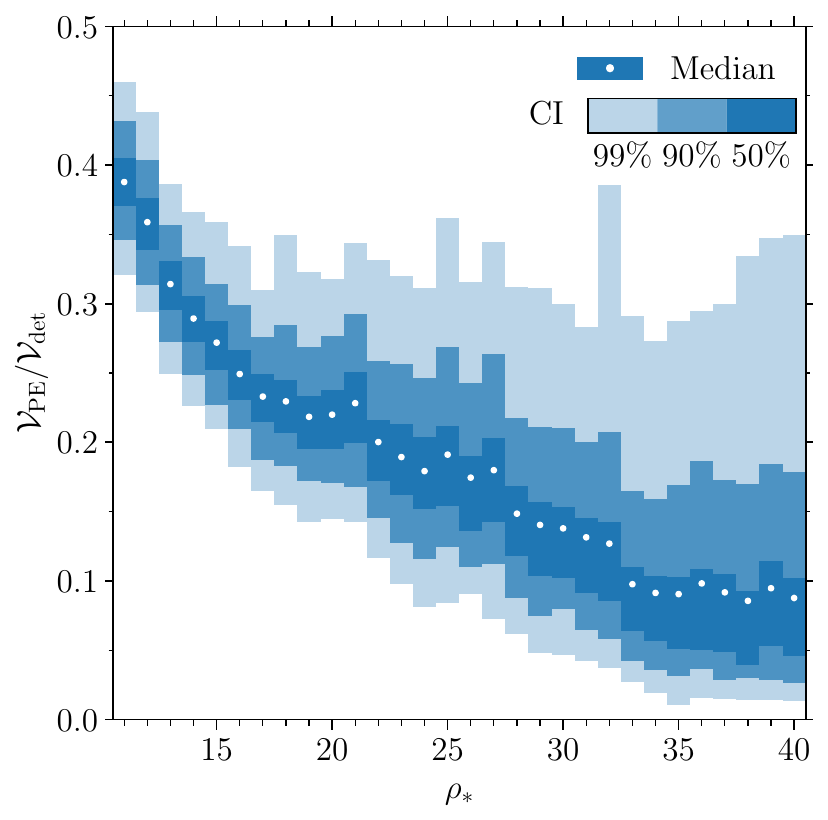}
    \caption{Monte Carlo variance \vpe{} from the sum of the variances when estimating the per-event likelihoods relative to the variance when estimating selection effects.
    White circles denote medians, while shades from darker to lighter denote the 99\%, 90\%, and 50\% \acp{CI}.
    }
    \label{fig:vpe-over-vdet-vs-rho}
\end{figure}

\begin{figure}[h]
    \centering
    \includegraphics[width=0.85\linewidth]{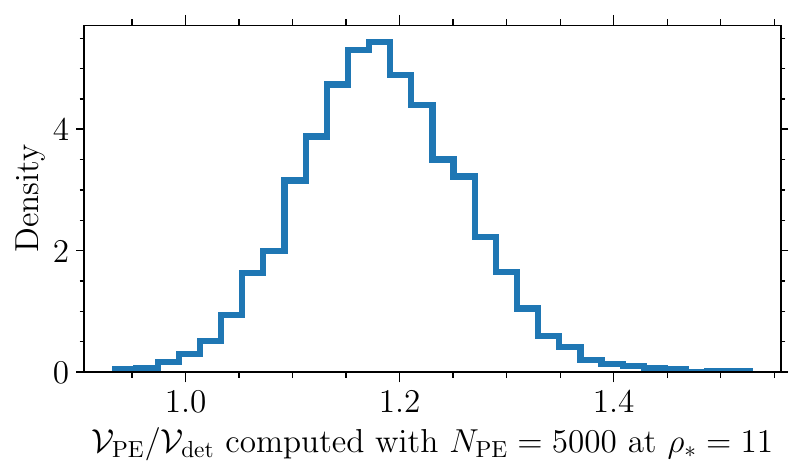}
    \caption{Distribution of $\vpe{} / \vdet{}$ over the posterior support when analyzing our largest mock catalog, at $\rho_* = 11$, using $\npe{} = 5000$ single-event posterior samples.
    For comparison, using $\npe{} \gtrsim 15000$ samples per-event (as in the analyses presented in this work) yields $\vpe{} / \vdet{} = 0.39^{+0.05}_{-0.04}$ (median and 90\% \ac{CI}) at $\rho_* = 11$ (cf. Fig.~\ref{fig:vpe-over-vdet-vs-rho}).
    }
    \label{fig:vpe-over-vdet-npe5000-rho11}
\end{figure}

In Sec.~\ref{sec:sys-unc}, we describe how the total Monte Carlo variance \lvar{} of the population log-likelihood estimator evolves with \ac{SNR} threshold.
We then focus on the number and computational cost of simulations required to accurately estimate $\ln \pdet{}$.
Here, we inspect the Monte Carlo variance \vpe{} carried by our estimate of the per-event likelihoods in Eq.~\eqref{eq:single-likelihood-given-lambda} relative to the variance \vdet{} carried by our estimate of $\ln \pdet{}$.

While previous work (e.g., Ref.~\cite{Talbot:2023pex}) indicates that selection effects may dominate Monte Carlo uncertainty for large catalogs, this trend may not hold when \nobs{} increases because the \ac{SNR} threshold decreases, as in our work.
As shown in App.~\ref{app:nmin-scaling}, $\vdet{} \propto \rho_*^{b - 2a}$,
assuming $\nobs \propto \rho_*^{-a}$ and $\ndet \propto \rho_*^{-b}$ for some powers $a$ and $b$.
Meanwhile, if each per-event likelihood estimate carries comparable variance, then \vpe{} may scale like $\vpe \propto \nobs \propto \rho_*^{-a}$.
Thus, we might expect that $\vpe{} / \vdet{} \propto \rho_*^{a - b}$, and for the astrophysical and draw distributions considered in this work we find $a - b \approx -0.04$.

In Fig.~\ref{fig:vpe-over-vdet-vs-rho} we show the \acp{CI} of $\vpe / \vdet$ over the posterior support obtained with each mock catalog selected with increasing \ac{SNR} thresholds $\rho_* \in \{ 11, \ldots 40 \}$.
At all thresholds, \vdet{} dominates the Monte Carlo uncertainty of the log-likelihood estimator.
Further, $\vpe{}$ contributes a larger share of the Monte Carlo uncertainty as the \ac{SNR} threshold is lowered and the size of the catalog increases.
However, as $\rho_*$ increases, $\vpe{} / \vdet{}$ decreases much faster than $\rho_*^{-0.04}$, with the median scaling approximately as $\rho_*^{-0.5}$.
This points to non-trivial dependence of both \vdet{} and \vpe{} on \ac{SNR} threshold.
For example, \vdet{} falls as the efficiency of importance sampling from $p(\theta | \ldraw)$ to $p(\theta | \ltrue)$ rises; cf. Eq.~\eqref{eq:vdet-vs-ndet}.
Since the population posterior density converges towards the true population with more data (i.e., lower \ac{SNR} threshold), we find that the median importance sampling efficiency improves (albeit with some scatter) as $\rho_*$ declines (cf. Fig.~\ref{fig:eff-vs-rho}), reducing the contribution of the selection term to the total variance in our largest mock catalogs.

Similarly, \vpe{} depends on both the number \npe{} and distribution of single-event posterior samples; in this work, we drew $> 15000$ posterior samples per event.
Although $\vpe{} < \vdet{}$ for all of the analyses considered in this work, this is not guaranteed for lower \npe{}.
In Fig.~\ref{fig:vpe-over-vdet-npe5000-rho11}, we show the distribution of $\vpe{} / \vdet{}$ over the posterior support at $\rho_* = 11$ if we used a factor of three fewer single-event posterior samples---corresponding to a factor of three fewer nested sampling live points when analyzing each event and a threefold decrease in compute time \cite{Ashton:2022grj}.
With $\npe{} = 5000$, the single-event variances dominate the total Monte Carlo variance of the population likelihood estimator for our largest mock catalog.

\section{Posterior error induced by Monte Carlo approximations} \label{app:ehat}

\begin{figure}[h!]
    \centering
    \includegraphics[width=0.9\linewidth]{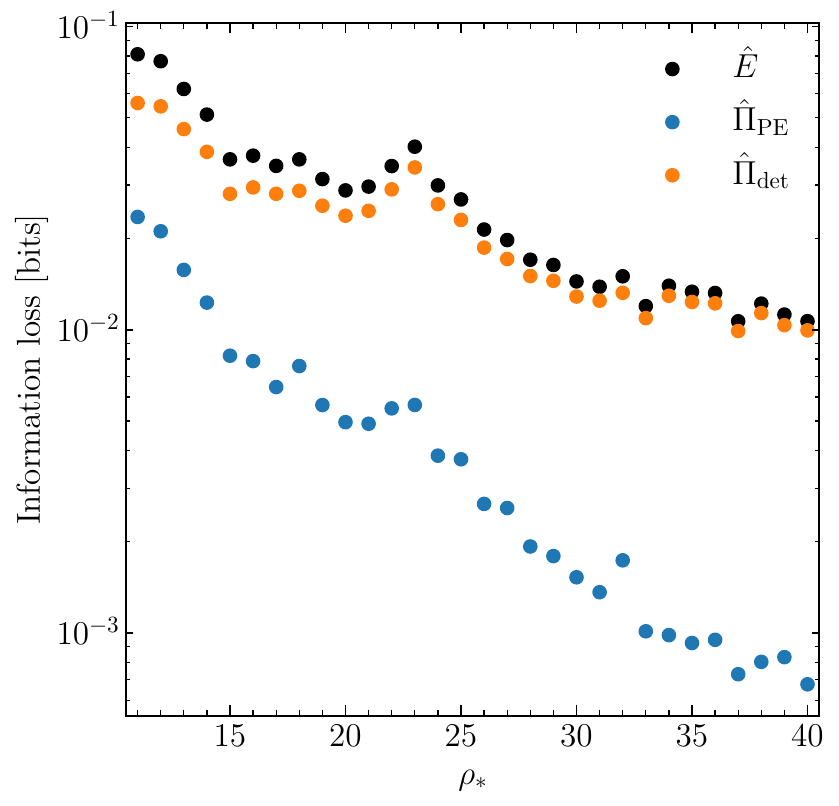}
    \caption{Posterior error $\hat{E}$ due to Monte Carlo approximation of the population likelihood and dominant contributions $\hat{\Pi}_{\rm PE}, \hat{\Pi}_{\rm det}$ to the error coming from Monte Carlo estimation of the per-event likelihoods and selection effects, respectively.}
    \label{fig:ehat}
\end{figure}

In this study we used the Monte Carlo variance of the likelihood estimator as a proxy for systematic bias in population inference.
However, Ref.~\cite{jack_math_tome} recently derived an estimator of the error in the \textit{posterior} induced by Monte Carlo approximation of the likelihood.
In Fig.~\ref{fig:ehat}, we show the posterior error estimate $\hat{E}$, plus the dominant contributions to $\hat{E}$, $\hat{\Pi}_{\rm PE}$ and $\hat{\Pi}_{\rm det}$, coming from Monte Carlo estimation of the per-event likelihoods and selection effects, respectively.
We find that all analyses satisfy a threshold of $\hat{E} \leq 0.2$\,bits recommended by Ref.~\cite{jack_math_tome}, indicating that no bias is present even in our analysis of all 1600 mock events.
Further, $\hat{E}$ decreases with increasing \ac{SNR} threshold, supporting our conclusion in Sec.~\ref{sec:sys-unc} that population analyses performed with higher significance threshold are less subject to systematic uncertainty due to misestimation of the population likelihood.
Finally, we note that uncertainty in the per-event likelihood estimates contributes a larger fraction of the total information loss at smaller \ac{SNR} thresholds, consistent with Fig.~\ref{fig:vpe-over-vdet-vs-rho}.

\clearpage

\bibliography{refs}

\end{document}